\documentclass{article}

\usepackage[utf8]{inputenc} % allow utf-8 input
\usepackage[T1]{fontenc}    % use 8-bit T1 fonts

\title{Multi-edge-type LDPC code design with G-EXIT charts for continuous-variable quantum key distribution}

\author{Hossein~Mani, Tobias~Gehring, Philipp~Grabenweger, Christoph~Pacher, and~Ulrik~Lund~Andersen 
	\thanks{H. Mani, T. Gehring and U. L. Andersen are from Center for Macroscopic Quantum States (bigQ), Department of Physics, Technical University of Denmark, $2800$ Kongens Lyngby, Denmark. Email: \{hosma, tobias.gehring, ulrik.andersen\}@fysik.dtu.dk.}
	\thanks{P.~Grabenweger and C. Pacher are with the Center for Digital Safety \& Security, AIT Austrian Institute of Technology GmbH, Giefinggasse $4$, $1210$ Vienna, Austria. Email: christoph.pacher@ait.ac.at.
		}% <-this % stops a space
}
\newcommand{\vecXA}{\vec{X}_A}
\newcommand{\vecXB}{\vec{X}_B}
\newcommand{\veca}{\vec{a}}

\usepackage{graphicx, amssymb, cite, amsmath, amsfonts, epsfig,cuted,lipsum,color,bbm, mathrsfs}
\usepackage{url}
\usepackage{hyperref}
\hypersetup{
    colorlinks=true,
    linkcolor=blue,
    filecolor=magenta,      
    urlcolor=cyan,
   pdftitle={MET-LDPC for CV-QKD},
    pdfauthor = {Hosma}
    pdfpagemode=FullScreen,
}

\usepackage{fullpage}
\usepackage{subfigure}
\usepackage[]{algorithm2e}
\usepackage{tikz} % Package for drawing
\usetikzlibrary{spy}
\usepackage{pgfplots}
\usetikzlibrary{decorations.pathreplacing,angles,quotes,shapes}
\graphicspath{{images/}{../images/}}

\normalsize

\begin{document}
	\maketitle
	\begin{abstract}
		Continuous-variable quantum key distribution utilizes an ensemble of coherent states of light to distribute secret encryption keys between two parties. One of the challenges is thereby the requirement of capacity approaching error correcting codes in the low signal-to-noise (SNR) regime (SNR  $< 0$ dB). Multilevel coding (MLC) combined with multistage decoding (MSD) can solve this challenge in combination with multi-edge-type low-density parity-check (MET-LDPC) codes which are ideal for low code rates in the low SNR regime due to degree-one variable nodes. However, the complexity of designing such highly efficient codes remains an open issue. Here, we introduce the concept of generalized extrinsic information transfer (G-EXIT) charts for MET-LDPC codes and demonstrate how this tool can be used to analyze their convergence behavior. We calculate the capacity for each level in the MLC-MSD scheme and use G-EXIT charts to exemplary find codes for some given rates which provide a better decoding threshold compared to previously reported codes. In comparison to the traditional density evolution method, G-EXIT charts offer a simple and fast asymptotic analysis tool for MET-LDPC codes.
	\end{abstract}
%\begin{IEEEkeywords}
%\end{IEEEkeywords}

% \IEEEpeerreviewmaketitle
\section{Introduction}\label{Sec:Introduction}
The security of today's asymmetric cryptography, e.g.\ the Rivest-Shamir-Adleman (RSA) protocol and the Diffie-Hellman key-exchange protocol, is based on mathematical complexity assumptions of basic problems like the discrete log problem and the factorization of large numbers~\cite{shor1999polynomial}. The advent of the quantum computer or even an unexpected algorithmic innovation can compromise their security with drastic consequences for the internet. One possible solution is quantum key distribution (QKD)~\cite{2009.QKD-review} which provides information theoretical secure cryptographic key exchange for two parties, Alice and Bob, based on the properties of quantum mechanics. In continuous-variable (CV) QKD~\cite{2009.QKD-review, Eleni_2015_CVQKD, 2018.Fabian} the transmitter, Alice, modulates (weak) coherent states and  the receiver, Bob, measures the amplitude and phase quadratures of the electromagnetic light field. The communication distance and the key generation rate is thereby limited by the performance of information reconciliation which is an important part in every QKD protocol to ensure that both parties generate the same cryptographic key. To achieve high transmission distances reverse reconciliation has to be applied, i.e.\ Alice has to reconcile on Bob's measurement results. The main challenge here is the design of capacity approaching error correction codes for very low signal-to-noise ratios (SNR). For instance in~\cite{jouguet2011long} an SNR of $-15.37$\,dB was reported for a transmission distance of $80$\,km and in~\cite{huang2016long} an SNR of $-16.198$\,dB for $100$\,km.      

Low-density parity-check (LDPC) codes in the multi-edge-type (MET) variant~\cite{richardson2002multi} can be used in combination with multilevel coding - multistage decoding (MLC-MSD) to perform capacity approaching error correction for QKD with low SNR~\cite{771140Multilevelcodes}. In~\cite{jouguet2011long}, the authors developed a code with rate $0.02$ with an efficiency of $96.9$\%, which  was also used in other works, e.g.\ \cite{milicevic2017key}. %The design of codes with low complexity is desirable to achieve encoder and decoder blocks with low implementation complexity.
Designing an efficient capacity achieving degree distribution (DD) has been investigated in many works and different analytical, as well, as numerical techniques have been introduced~\cite{CapAch_Shokrollahi2002, CapAch_Urbanke2005, CapAch_Saeedi2010}.

Traditionally, code design for LDPC codes is a time consuming process using the density evolution algorithm~\cite{2004.Density.Evolution}. In each iteration of the density evolution algorithm a vector of real values representing the densities has to be updated which is computationally expensive. Due to this complexity many approximation methods for density evolution have been developed, for instance, Gaussian approximation and Extrinsic Information Transfer (EXIT) charts~\cite{CapAch_Saeedi2010,Ardakani2004}. Nowadays, for a variety of binary memory-less channels these asymptotic analysis tools are used for the optimization of the degree distribution. Specifically, for the binary erasure channel (BEC), the design of capacity achieving codes can be carried out by matching the two curves of EXIT functions related to the variable node and check node degree distributions due to the area theorem~\cite{CapAch_Shokrollahi2002}. In other binary input-output symmetric memory-less channels the Generalized area theorem and  generalized EXIT (G-EXIT) charts can be used for the optimization problem~\cite{Measson2009GAT}.

Here, we introduce G-EXIT charts for MET-LDPC codes to provide a practical tool for their design and optimization. We use our tool to design new highly efficient codes with rates $0.02,\,0.05,\,$ and $0.1$. 

In general for a given input distribution we calculate the Shannon capacity for each level in the MLC-MSD scheme and present highly efficient MET-LDPC codes for various SNRs. 

The organization of the remainder of this paper is as follows. In Section\,\ref{sec:system model}, we explain the system model for the reconciliation of the secret key rate and calculate the designed capacity rate for each level for a given input distribution. Section\,\ref{sec:ME-LDPC} briefly reviews the basic concepts of MET-LDPC codes and the extension of density evolution to these codes. Then in Section\,\ref{sec:G-EXIT}, we introduce the concept of G-EXIT charts for MET-LDPC codes. Simulation results are presented in Section\,\ref{sec:Code Design}, where we show how to use a G-EXIT chart for designing MET-LDPC codes. Finally, Section\,\ref{sec:Conclusion} concludes the paper.

%%%%%%%%%%%%%%%%%%%%%%%%%%%%%%%%%%%%%%%%%%% 
%%%%%%%%%%%%%%%%%%%%%%%%%%%%%%%%%%%%%%%%%%%  SEC II
%%%%%%%%%%%%%%%%%%%%%%%%%%%%%%%%%%%%%%%%%%% 
\section{System Model}\label{sec:system model}
\subsection{Source Coding with side information and equivalent channel coding model}
Information reconciliation is a method by which two parties that each possess a sequence of numbers agree on a \emph{common} sequence of bits by exchanging one or more messages. Mathematically speaking, in CV-QKD with Gaussian modulation the two sequences of numbers are joint instances of a bivariate random variable that follows a bivariate normal distribution. Physically, these sequences are obtained by one party generating coherent states in the quadrature phase space and the other party measuring them. In other words, in QKD two parties share correlated random variables and wish to agree on a common bit sequence. 
However, imperfect correlations introduced by the inherent shot noise of coherent states and noise in the quantum channel and the receiver, give rise to discrepancies in the two sequences of numbers which have to be corrected by exchanging additional information.

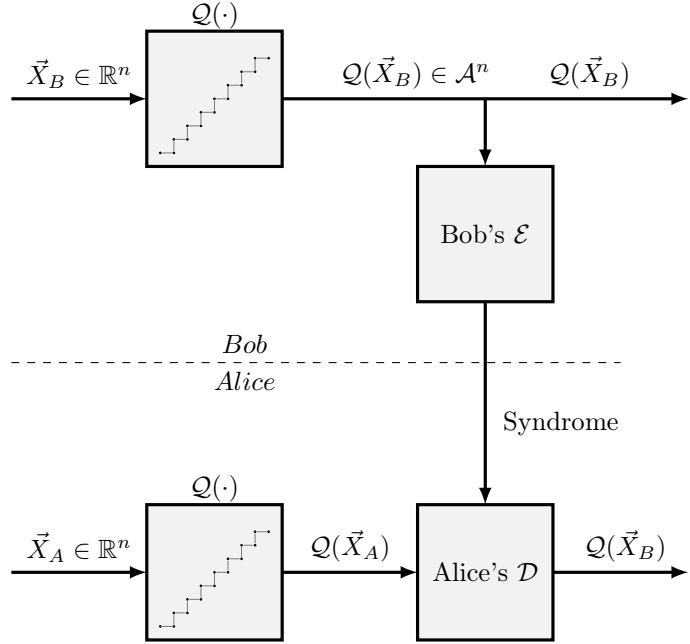
\begin{figure}[t]
\begin{center}
%\resizebox{0.8\hsize}{!}{
		\begin{tikzpicture}[scale=0.90]
		\draw [black, very thick, >=latex] [->] (-4,2) -- (-2,2) node[above, pos=.5] { $\vecXB \in \mathbb{R}^n$};
		\draw[black, very thick, fill=gray!10!white] (-2,1) rectangle ++(2,2) node[above, pos=.5, yshift=.8 cm] {$\mathcal{Q}(\cdot)$};
		\def\mypoints{(-1.8, 1.2),  (-1.6,1.2), (-1.6,1.4), (-1.4,1.4), (-1.4,1.6),
  (-1.2,1.6), (-1.2,1.8), (-1,1.8), (-1,2), (-.8,2), (-.8,2.2) , (-.6,2.2),   (-.6,2.4), (-.4,2.4), (-.4,2.6), (-.2,2.6)};
        \path
        \foreach \x [count=\xi] in \mypoints {
            \x node[circle, fill, inner sep=sqrt(2)*0.00025cm] (node\xi) {}
        }
        \foreach \x [count=\xi, remember=\xi-1 as \xiprev] in \mypoints {
            \ifnum\xi>1 %
            (node\xiprev) edge[>=latex, black!50!white] (node\xi)
            \fi
        };
		\draw [black, very thick, >=latex] [->] (0,2) -- (6,2) node[above, pos=.5] { $\mathcal{Q}(\vecXB) \in \mathcal{A}^n\,\,\,\,\,\,\,\quad \mathcal{Q}(\vecXB) $};
		\draw [black, very thick, >=latex] [->] (3,2) -- (3,1); % Vertical line
		\draw[black, very thick, fill=gray!10!white] (2,-1) rectangle ++(2,2) node[pos=.5] {Bob's $\mathcal{E}$};
		\draw [black, dashed, >=latex] [-] (-4,-1.9) -- (5,-1.9) node[below, xshift=-5 cm] { $Alice$} node[above, xshift=-5 cm] { $Bob$};
		\draw [black, very thick, >=latex] [->] (3,-1) -- (3,-4) node[right, xshift = .1 cm, pos = 0.6] {Syndrome}; % Vertical line
			\draw[black, very thick, fill=gray!10!white] (2,-6) rectangle ++(2,2) node[pos=.5] {Alice's $\mathcal{D}$};
		\draw [black, very thick, >=latex] [->] (-4,-5) -- (-2,-5) node[above, pos=.5] { $\vecXA \in \mathbb{R}^n$};
		\draw [black, very thick, >=latex] [->] (0,-5) -- (2,-5)node[above, pos=.5] { $\mathcal{Q}(\vecXA)$};
		\draw[black, very thick, fill=gray!10!white] (-2,-6) rectangle ++(2,2) node[above, pos=.5, yshift=.8 cm] {$\mathcal{Q}(\cdot)$};
		\def\mypoints{(-1.8, -5.8),  (-1.6,-5.8), (-1.6,-5.6), (-1.4,-5.6), (-1.4,-5.4),
  (-1.2,-5.4), (-1.2,-5.2), (-1,-5.2), (-1,-5.0), (-.8,-5), (-.8,-4.8) , (-.6,-4.8),   (-.6,-4.6), (-.4,-4.6), (-.4,-4.4), (-.2,-4.4)};
        \path
        \foreach \x [count=\xi] in \mypoints {
            \x node[circle, fill, inner sep=sqrt(2)*0.00025cm] (node\xi) {}
        }
        \foreach \x [count=\xi, remember=\xi-1 as \xiprev] in \mypoints {
            \ifnum\xi>1 %
            (node\xiprev) edge[>=latex, black!50!white] (node\xi)
            \fi
        };
		\draw [black, very thick, >=latex] [->] (4,-5) -- (6,-5) node[above, pos=.5] { $\,\,\mathcal{Q}(\vecXB) $};
		\end{tikzpicture}
%		}
	\end{center}
	\caption{Correlated source coding configuration. Correlated information sequences $\vecXB= (X_{B,0}, X_{B,1}, \ldots X_{B,n-1})$ and $\vecXA=(X_{A,0}, X_{A,1}, \ldots X_{A,n-1})$ are generated by a pair of continuous random variables $X_A, X_B$ from a given bivariate distribution $p_{X_AX_B}(x_A,x_B)$.}\label{fig:sourceCoding}
\end{figure}

In reverse reconciliation, which is the focus of this paper, Alice reconciles her values to match Bob's values. The reconciliation process can be fully described as a conventional communication theory problem. This problem was first addressed in~\cite{1055037Wolf} as source coding with side information: Let Alice and Bob have access to two correlated information sources $X_A$ and $X_B$ which follow a joint probability distribution $p_{X_AX_B}(x_A,x_B)$. The two parties wish to distill a common binary string from blocks of length $n$, $\vecXA=\left(X_{A,i}\right)_{i=0}^{n-1}$, $\vecXB=\left(X_{B,i}\right)_{i=0}^{n-1}$, by exchanging information as shown in Fig.~\ref{fig:sourceCoding}. In this configuration Bob sends to Alice a compressed version of his block of discretized (quantized) symbols $\mathcal{Q}(\vecXB) = \left(\mathcal{Q}(X_{B,i})\right)_{i=0}^{n-1}$ and knows that Alice has access to the side information $\vecXA$.
Furthermore, it is convenient to generate the reconciliation messages as syndromes using linear codes with performance close to the Shannon capacity~\cite{1633793Bloch,design_BIAWGN_2}. In this case the parity-check matrix of the error correction code can be used to generate the syndrome for the reconciliation problem. Thus an equivalent channel coding problem can be solved instead of the above mentioned source coding with side information. In the following we use an equivalent MLC-MSD scheme to design asymptotically optimal encoder-decoder blocks. 
%%%% =====================================================
%%%% =====================================================
%%%% =====================================================
\subsection{Slice Reconciliation based on MLC-MSD}
Slice reconciliation using error correction codes can be described in two steps. The first step is discretization which transforms the continuous Gaussian source $X_B$ into an $m$ bit source $\mathcal{Q}(X_B)$ with its binary representation vectors $(X_B^{m-1}, \ldots, X_B^1, X_B^0)$. There is an inherent information loss due to the discretization process of the source. The second step can be modeled as source coding with side information on the MLC-MSD scheme. In reverse reconciliation, as presented in Fig.~\ref{fig:sourceCoding}, Bob sends an encoding (compressed version) of $\mathcal{Q}(\vecXB)$ to Alice, such that she can infer $\mathcal{Q}(\vecXB)$ with high probability using her own source $\vecXA$ as side information.

%\textbf{Efficiency}
Let us define the efficiency $\beta$ as
\begin{equation}\label{eq:efficiency}
\beta =\dfrac{H(\mathcal{Q}(X_B))-R^s}{I(X_B;X_A)}\ , 
\end{equation}
where  $I(X_B;X_A)$ is the mutual information and $H(\mathcal{Q}(X_B))-R^s$ is the net shared information between two parties, resp.\ per symbol, \cite{jouguet2014high,Cardinal_2003} with $H(\cdot)$ the Shannon entropy and $R^s$ the source coding rate.
Thus the practical efficiency of the reconciliation depends on the ability to design very good discretizers and highly efficient compression codes with minimum possible source coding rate. Note, that Slepian and Wolf \cite{1055037Wolf} have shown that $H(Y|Z)$ is the lower bound to the source coding rate when decoding $Y$ given side information $Z$. Therefore, $R\textsuperscript{s} \geq H(\mathcal{Q}(X_B)|X_A)$. 

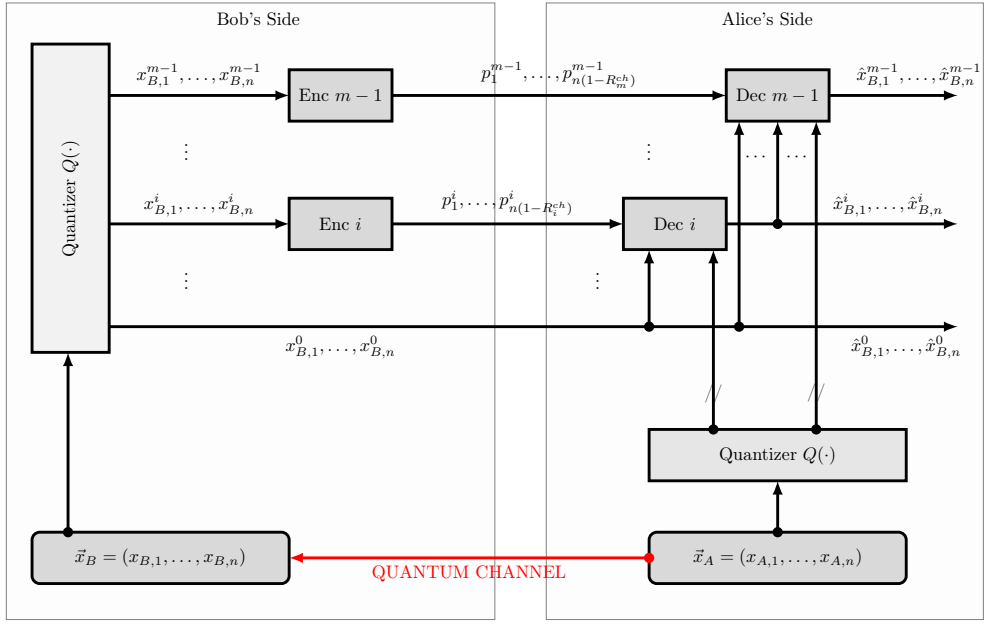
\begin{figure*}[t]
	\begin{center}
% 		{%\resizebox{14cm}{9 cm}
		\resizebox{0.8\hsize}{!}{
			\begin{tikzpicture}[scale=1.0, 
			dot/.style    = {anchor=base,fill,circle,inner sep=2pt},
			block_1/.style  = {draw, rectangle, black, ultra thick, fill=gray!30!white, minimum width = 0.75cm, minimum height = 0.75cm},]

			\draw[gray, ultra thin, fill=gray!2!white] (0.0,-8.2) rectangle ++(9.5,12) node at (4.9,3.5) [black]{Bob's Side};
			\draw[gray, ultra thin, fill=gray!2!white] (10.5,-8.2) rectangle ++(8.5,12)  node at (14.8,3.5) [black]{Alice's Side};
			
			%% QUANTUM CHANNEL AT THE BOTOM
			\draw[black, ultra thick, fill=gray!30!white, rounded corners=1ex] (0.5,-7.5) rectangle ++(5,1) node[pos=.5] {$\vec{x}_B = (x_{B,1}, \dots, x_{B,n})$};
			\draw[black, ultra thick, fill=gray!30!white, rounded corners=1ex] (12.5,-7.5) rectangle ++(5,1) node[pos=.5] {$\vec{x}_A = (x_{A,1}, \dots, x_{A,n})$};
			\draw[black, ultra thick, fill=black!10!white] (12.5,-5.50) rectangle ++(5,1) node[pos=.5] {Quantizer $Q(\cdot)$};
			\draw [black, ultra thick, >=latex] [->] (15,-6.5) coordinate[dot] -- (15,-5.5) ;
			\draw [red, ultra thick, >=latex] [->] (12.5,-7) coordinate[dot] -- (5.5,-7) node[pos=0.5, below]{QUANTUM CHANNEL};

			% CLASSICAL CHANNEL
			\draw[black, ultra thick, fill=gray!10!white] (.5,-3) rectangle ++(1.5,6) node[rotate=90,pos=0.5]{Quantizer $Q(\cdot)$};
			\draw [black, ultra thick, >=latex] [->] (1.2,-6.5) coordinate[dot] -- (1.2,-3) ;
			% ENCODER SIDE:
			\draw [black, ultra thick, >=latex] [->] (2,2) -- (5.5,2) node[above, pos=.5] { $x_{B,1}^{m-1},\ldots, x_{B,n}^{m-1}$};
			\draw [black, ultra thick, >=latex] [->] (2,-.5) -- (5.5,-.5) node[above, pos=.5] { $x_{B,1}^i,\ldots, x_{B,n}^i$};
			\draw [black, ultra thick, >=latex] [->] (2,-2.5) -- (18.5,-2.5);
			\draw[black, ultra thick, fill=gray!30!white] (5.5,1.5) rectangle ++(2,1) node[pos=.5] {Enc $m-1$};
			\draw[black, ultra thick, fill=gray!30!white] (5.5,-1) rectangle ++(2,1) node[pos=.5] {Enc $i$};
			\draw [black, ultra thick, >=latex] [->] (7.5,2) -- (14,2) node[above, pos=.5] { $p_1^{m-1},\ldots, p_{n(1-R^{ch}_m)}^{m-1}$};
			\draw [black, ultra thick, >=latex] [->] (7.5,-.5) -- (12,-.5) node[above, pos=.5] { $p_1^i,\ldots, p_{n(1-R^{ch}_i)}^i$};
			%% DECODER SIDE
			\draw[black, ultra thick, fill=gray!30!white] (12,-1) rectangle ++(2,1) node[pos=.5] {Dec $i$};
			\draw[black, ultra thick, fill=gray!30!white] (14,1.5) rectangle ++(2,1) node[pos=.5] {Dec $m-1$};
			\draw [black, ultra thick, >=latex] [->] (16,2) -- (18.5,2) node [pos = 0.7, above] {$\hat{x}_{B,1}^{m-1},\ldots, \hat{x}_{B,n}^{m-1}$};
			\draw [black, ultra thick, >=latex] [->] (14,-.5) -- (18.5,-.5) node [pos=.7, above] {$\hat{x}_{B,1}^i,\ldots, \hat{x}_{B,n}^i$};
			%% VERTICAL LINES AND DOTS
			\draw [black, ultra thick, >=latex] [->] (12.5,-2.5) coordinate[dot] -- (12.5,-1) ;
			\draw [black, ultra thick,  >=latex] [->] (13.75,-4.5) coordinate[dot] -- (13.75,-1) node [pos=.2]{//};
			\draw [black, ultra thick, >=latex] [->] (14.25,-2.5) coordinate[dot] -- (14.25,1.5) ;
			\draw [black, ultra thick, >=latex] [->] (15,-.5) coordinate[dot] -- (15,1.5) ;
			\draw [black, ultra thick, >=latex] [->] (15.75,-4.5) coordinate[dot] -- (15.75,1.5) node [pos=.12]{//};
			\node at (17.5,-2.9) {$\hat{x}_{B,1}^0,\ldots, \hat{x}_{B,n}^0$};
			\node at (6.5,-2.9) {$x_{B,1}^0,\ldots, x_{B,n}^0$};
			\node at (3.5,-1.5) {$\vdots$};
			\node at (11.5,-1.5) {$\vdots$};
			\node at (3.5,1.) {$\vdots$};
			\node at (12.5,1.) {$\vdots$};
			\node at (14.6,0.8) {$\cdots$};
			\node at (15.4,0.8) {$\cdots$};
			\end{tikzpicture}
		}
	\end{center}
	\caption{The MLC-MSD scenario for the reverse reconciliation. First the input source is quantized into an $m$-bit source. Then each of the $m$ sources is encoded and sent to Alice. The decoder has the side information from its own source and with the $m$ encoded sources produces an estimate of the quantized source. Usually we transmit the least significant bits directly to the channel.}\label{fig:ChannelCodingRR}
\end{figure*}

A more detailed schematic representation for MLC-MSD scheme is presented on Fig.~\ref{fig:ChannelCodingRR}, where Bob encode his data on $m$ different individual levels. Let us denote by $R^s_i$ the corresponding source coding rate for each sub-level $i$ in the MLC-MSD scheme. Then according to Fig.~\ref{fig:ChannelCodingRR}, and using again the Slepian-Wolf theorem, the $R_i^s$ are lower bounded by the conditional entropy of the $i$-th bit of $X_B$, given side information $X_A$ and all the lower bits of $X_B$:
\begin{equation}\label{eq:source_coding_SW}
R^s_i \geq H(X_{B}^i|X_A, X_{B}^{i-1},\ldots, X_{B}^0)\ .
\end{equation} 
The total source coding rate is given by summing over the individual source code rates $R^s_i$:
\begin{eqnarray}\label{eq:sorceTochannelRate}
\nonumber R^s = \sum_{i=0}^{m-1}R^s_i,
\end{eqnarray}
and we resemble the Slepian-Wolf theorem again:
\begin{eqnarray}\label{eq:sorceTochannelRate2}
\nonumber R^s &\geq \sum\limits_{i=0}^{m-1}H(X_{B}^i|X_A,X_{B}^{i-1},\ldots, X_{B}^0) = H(\mathcal{Q}(X_B)|X_A).
\end{eqnarray}

The detailed block diagram for the MLC-MSD scheme for reverse reconciliation is depicted in Fig.~\ref{fig:ChannelCodingRR}.
We consider a quantization scheme with $M = 2^m$, $m > 1$, signal points in a $D$-dimensional real signal space, with signal points taken from the signal set $\mathbf{S} = \{\veca_0, \veca_1, \ldots, \veca_{M-1}\}$ with probabilities $\Pr\{\veca_k\}$. Each signal point has its equivalent binary form defined by a bijective mapping $\veca = \mathcal{M}(\vec{x})$ of binary representation vectors $\vec{x} = (x_B^{m-1}, \ldots, x_B^0)$ to signal points $\veca \in \mathbf{S}$. Two well defined mappings are binary and Gray mapping. As an example for $m=3$ levels, in one-dimensional signal space ($D = 1$), the $M = 2^3$ signal points are taken from $\mathbf{S} = \{-7, -5, -3, -1, +1, +3, +5, +7\}$. 
% amplitude-shift keying (ASK) modulation with
Fixing the values of co-ordinates $i$ to $0$, i.e.\ $x_B^{i},\dots,x_B^0$, we obtain subsets of the signal set $\mathbf{S}$:
\begin{align}\label{eq:mapping}
\nonumber \mathbf{S}(x_B^{i}, \ldots, x_B^0)&=\\ 
\nonumber \{\veca = \mathcal{M}(\vec{x}) ~|~& \vec{x} = (b^{m-1}, \ldots, b^{i+1}, x_B^{i}, \ldots, x_B^0), b^j \in\{0,1\},~  j = i+1, \ldots, m-1\}. 
\end{align}
For more details about set partitioning and mapping see~\cite{771140Multilevelcodes}. For example, for the above mentioned constellation points with $M = 8$ with binary partitioning we have:
\begin{align}
\nonumber &\mathbf{S}(x_B^0 = 0)  = \left\{\veca = \mathcal{M}(\vec{x})| \vec{x}= \{000, 010, 100, 110\} \right\}~= \{-7,  -3,+1, +5\},\\
\nonumber &\mathbf{S}(x_B^1x_B^0 = 10)  = \{\veca = \mathcal{M}(\vec{x})|\vec{x}= \{010, 110\} \} ~= \{-3,+5\},\\
\nonumber  &\mathbf{S}(x_B^2x_B^1x_B^0 = 010) = \{\veca = \mathcal{M}(\vec{x})| \vec{x} = \{010\} \} = \{-3\}.
\end{align}
%%%%%%%%%%%%%%%%%%%%%%%%%%%%%%%%%%%%%%%%%%%%%%
\subsection{Classical statistical representation}
In the following we assume that $X_A$ and $X_B$ are jointly distributed according to a bivariate normal distribution with zero mean. The bi-variate normal distribution can be described by
\begin{equation}\label{eq:jointly_gaussian_dist}
%\resizebox{.89\hsize}{!}{$%
f_{X_B, X_A}(x_B,x_A) = \dfrac{1}{2\pi\sqrt{|\Sigma|}}\exp\left(-\frac{1}{2}(x_A,x_B)\Sigma^{-1}(x_A, x_B)^T\right)
%$}
\end{equation}
with the covariance matrix
\begin{equation}\label{eq:covariance_mat}
\Sigma = \begin{bmatrix}
\sigma^2_A       & \rho\,\sigma_A\,\sigma_B\\
\rho\,\sigma_A\,\sigma_B & \sigma^2_B
\end{bmatrix}\ ,
\end{equation}
where
\begin{equation}\label{eq:correlation_coef}
\rho = \dfrac{\mathbb{E}\{X_AX_B\}}{\sigma_A\sigma_B}\ ,
\end{equation}
is the correlation coefficient between $X_A$ and $X_B$ where $\sigma_A$ and $\sigma_B$ denotes their standard deviations, respectively. We empirically estimate the covariance matrix during the parameter estimation phase of the quantum key distribution protocol. We then normalize Alice's and Bob's data by dividing by their respective standard deviation, i.e. 
\begin{align*}
x^j_A &\rightarrow x^j_A / {\sigma}_A\ , \\
x^j_B &\rightarrow x^j_B / {\sigma}_B\ , 
\end{align*}
such that
\begin{equation}
\Sigma \rightarrow \Sigma =
\begin{bmatrix}
1       & \rho \\
\rho & 1
\end{bmatrix}\ .
\end{equation}
The conditional probability distribution describing Alice's outcome conditioned on Bob's is given by
\begin{equation}\label{eq:conditional_pdf_cont_Alice_givenBob}
f_{X_A|X_B}(x_A|x_B) = \mathcal{N}(\rho x_B, (1-\rho^2))\  .
\end{equation}
%%%%%%%%%%%%%%%%%%%%%%%%%%%%%%%%%%%%
\subsection{Quantization Effect}
As we discussed on Section \ref{sec:system model}-B, let us denote the quantized version of $X_B$ by $Q(X_B)$ with its binary equivalent vector $(X_B^{m-1}, \ldots, X_B^1, X_B^0)$. Considering a fixed step size $\delta$ for discretization the entropy of the quantized source can be approximated by $$H(Q(X_B))   \approx  h(X_B) - \log_2\delta\ ,$$ where $h(X_B)$ is the differential entropy defined for continuous variable $X_B$. A similar quantization can be applied on the Alice's side to get $Q(X_A)$. This also holds for the conditional entropy, that is $$H(Q(X_B)|Q(X_A))   \approx  h(X_B|X_A) - \log_2\delta\ .$$
If $m$ is large enough, and thus $\delta$ is small, we can approximate
$$I(Q(X_B);Q(X_A)) \approx I(Q(X_B);X_A) \approx I(X_B;X_A)\ ,$$
where the equality holds when $\delta \rightarrow 0.$ 

%%%%%%%%%%%%%%%%%%%%%%%%%%%%%%%%%%%%
\subsection{Individual source coding rates}
The individual conditional mutual information for each level is defined as 
\begin{align}\label{eq:I_i_RR}
\nonumber I_i &= I(X_A; X_{B}^i|X_{B}^{i-1},\ldots,X_{B}^0) \\ 
&= H(X_{B}^i|X_{B}^{i-1},\ldots,X_{B}^0)-H(X_{B}^i|X_A, X_{B}^{i-1},\ldots,X_{B}^0)\ .
\end{align}
Thus using Eq.\ \eqref{eq:source_coding_SW} we obtain
\begin{equation}\label{eq:R_i_RR_simple}
R^s_i \ge H(X_{B}^i|X_{B}^{i-1},\ldots,X_{B}^0) - I_i\ .
\end{equation}
Equation \eqref{eq:R_i_RR_simple} offers an analytical way to calculate the individual source rates and the only non-trivial thing we need to do is to calculate the quantity $I_i$. We will now apply the MLC-MSD approach to calculate individual conditional mutual information. It is also noteworthy to mention that, using the chain rule we can always describe the total mutual information as a summation of conditional mutual information for individual levels 
\begin{align}\label{eq:I_i_vs_I_tot}
\nonumber I(X_A; Q(X_B)) &=~ I(X_A;X_{B}^{m-1},\ldots,X_{B}^0)\\
\nonumber &=~I(X_A;X_{B}^0) + I(X_A; X_{B}^1|X_{B}^0) + \ldots+I(X_A; X_{B}^i|X_{B}^{i-1},\ldots,X_{B}^0)+ \\
&~~~\ldots+I(X_A; X_{B}^{m-1}|X_{B}^{m-2},\ldots,X_{B}^0)\ ,
\end{align}
which motivates our definition of the individual conditional mutual information in \eqref{eq:I_i_RR}. According to \cite{771140Multilevelcodes}, we can expand the $I_i$  as follows:
\begin{align}\label{eq:calc_I_RR_i}
\nonumber I_i &=\,I(X_A; X_{B}^i|X_{B}^{i-1},\ldots,X_{B}^0)\\ 
 &=\, I(X_A; X_{B}^{m-1}, \ldots, X_{B}^i|X_{B}^{i-1},\ldots,X_{B}^0)-I(X_A; X_{B}^{m-1}, \ldots, X_{B}^{i+1}|X_{B}^{i},\ldots,X_{B}^0)\ , 
\end{align}
where then each term on the right hand side can be calculated separately. More precisely, 
  \begin{align*}
    I(X_A; X_{B}^{m-1},\ldots,X_{B}^{i}&|X_{B}^{i-1},\ldots,X_{B}^0)= \\
    & \mathbb{E}_{x^{i-1}_{B},\ldots,x_{B}^0\,\in\,\{0,1\}^i}\left\{I(X_A; X_{B}^{m-1}, \ldots, X_{B}^{i}|x_{B}^{i-1},\ldots,x_{B}^0)\right\}\ ,
\end{align*}

where it can be calculated by averaging over all possible combinations of $x_{B}^{i-1},\ldots,x_{B}^0$. Finally, according to \cite{771140Multilevelcodes}, the full characterization of $I_i$, requires a set of  probability density functions (PDF)s $\mbox{\textbf{\textit{f}}}_{X_A|X_B^i}(x_A|x_B^i)$ which can be defined as:
\begin{align*}
   \mbox{\textbf{\textit{f}}}_{X_A|X_B^i}(x_A|x_B^i) = \left\{f_{X_A|X_B}\left(x_A|x_B^i, x_{B}^{i-1},\ldots,x_{B}^0\right)|\left(x_{B}^{i-1},\ldots,x_{B}^0\right) \in\{0,1\}^i\right\}\ ,
\end{align*}
where,
\begin{align*}
   f_{X_A|X_B}(x_A|x_B^i, x_{B}^{i-1},\ldots,x_{B}^0) = \mathbb{E}_{b\,\in\, \mathbf{S}(x_B^{i}, \ldots, x_B^0)}\left\{f_{X_A|X_B}(x_A|b)\right\},
\end{align*}
where the signal point $b$ is taken from the subset $\mathbf{S}(x_B^0\ldots\,x_B^{i})$ and $f_{X_A|X_B}(x_A|b)$ is the equivalent conditional quantized PDF of the continuous conditional PDF presented on \eqref{eq:conditional_pdf_cont_Alice_givenBob}.

\begin{figure}[t]
	\begin{center}
			\begin{tikzpicture}[spy using outlines={circle,gray,magnification=6,size=2cm, connect spies}]
		\node {\includegraphics[width=0.6\textwidth,keepaspectratio]{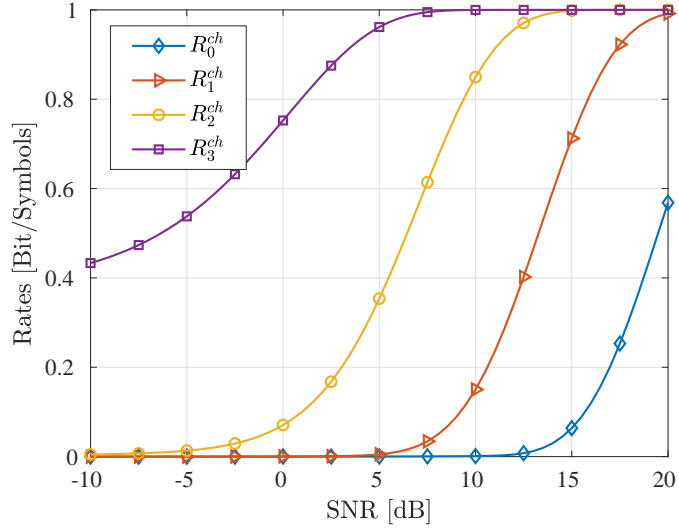}};
		%\spy on (-2.,-2.1) in node [left] at (2.5,1.3); %(-2.7,-3.2) in node [left] at (3,2);
		\end{tikzpicture}
		%[width = 8.0cm, height = 7 cm]
	\end{center}\caption{The equivalent channel coding rates $(R_i^{ch} = 1-R_i^s)$ for $4$-levels with binary partitioning versus SNR. Here we assumed that Alice has a continuous Gaussian distribution with variance equal to $1$. Also a fixed step size $\delta = 0.32$ is assumed for quantization of normalized continuous variables.}\label{fig:individualratesDGD_cor}
\end{figure}

\begin{figure}[!ht]
	\begin{center}
			\begin{tikzpicture}%[spy using outlines={circle,gray,magnification=6,size=2cm, connect spies}]
		\node {\includegraphics[width=0.6\textwidth,keepaspectratio]{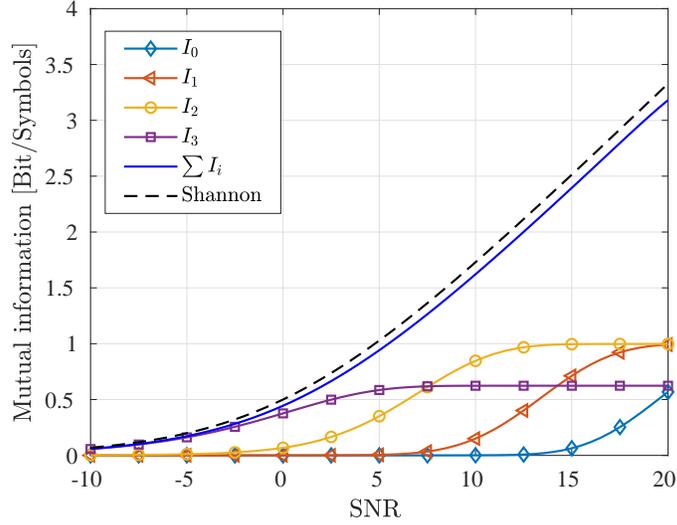}};
		%\spy on (-2.,-2.1) in node [left] at (2.5,1.3); %(-2.7,-3.2) in node [left] at (3,2);
		\end{tikzpicture}
		%[width = 8.0cm, height = 7 cm]
	\end{center}\caption{The individual conditional mutual information for the $4$-levels with binary partitioning versus SNR. It is assumed that Alice has a Gaussian distribution with variance $1$ and both Alice and Bob use $4$ level quantizers.}\label{fig:individualratesDGD}
\end{figure}

As an example, we present simulation results for individual rates for $16$-bins, when the Bob's quantized data has a discrete Gaussian PDF. 
The results for individual channel coding rates are presented in Fig.~\ref{fig:individualratesDGD_cor}. The equivalent channel coding rate is equal to $R_i^{ch} = 1-R_i^s$. 
In Fig.~\ref{fig:individualratesDGD}, we calculated the conditional mutual information $I_i$ of the individual $i$-th channel as a function of the SNR.  In addition, the summation of the individual conditional mutual information is compared with the Shannon capacity of the AWGN channel, which shows the quantization effect. According to \eqref{eq:I_i_RR} it is clear that 
\begin{align*}\label{eq:Capacity_and_SlepianWolf}
    \sum_{i=0}^{m-1} I_i = I(\mathcal{Q}(X_B);X_A) \leq I(X_B;X_A).
\end{align*}

%%%%%%%%%%%%%%%%%%%%%%%%%%%%%%%%%% 
%%%%%%%%%%%%%%%%%%%%%%%%%%%%%%%%%%  SEC III
%%%%%%%%%%%%%%%%%%%%%%%%%%%%%%%%%%
\section{MET-LDPC codes}\label{sec:ME-LDPC}
\subsection{MET-LDPC code ensemble}
\par Multi-edge-type (MET) LDPC codes  are a generalization of the concept of irregular LDPC codes~\cite{richardson2008modern, richardson2002multi}. These codes provide improvements in performance and complexity by giving more flexibility over different edge types. In this structure each node is characterized by the number of connections (sockets) to edges of each edge-type. It is noteworthy to mention that an irregular LDPC code is a single-edge-type LDPC (SET-LDPC) code. Using MET-LDPC codes we are able to design capacity achieving codes without using variable nodes with very high degree which provides a less complex implementation. Also it exploits the advantage of using degree one variable nodes, which are very useful for designing LDPC codes at low rate and low SNR\cite{richardson2002multi}. It is important to recall that in the case of SET-LDPC codes the minimum variable node degree is $2$.
\par A graph ensemble is specified through two multi-variable-polynomials, one associated with variable nodes and the other associated with check nodes. We denote these multi-variable-polynomials by
\begin{equation}\label{eq:dd4MET-LDPC}
\nu(\mathbf{r,x}) = \sum \nu_{\mathbf{bd}} \mathbf{r^b} \mathbf{x^d}, \mbox{        	} \mu(\mathbf{x}) = \sum \mu_{\mathbf{d}}\mathbf{x^d}
\end{equation}
respectively, where in Eq.~\eqref{eq:dd4MET-LDPC} we define the vectors $\mathbf{b}, \mathbf{d}, \mathbf{r}, \mathbf{x}$ and the coefficients $\nu_{\mathbf{bd}}$ and $\mu_{\mathbf{d}}$ as follows. Let $n_e$ denote the number of edge types and $n_r$ denote the number of different channels over which the code-word bits can be transmitted. To represent the structure of the graph we introduce the following \textit{node-perspective} multi-variable-polynomial representation. We thereby interpret degrees as exponents. Let $\mathbf{d} := (d_1, \ldots, d_{n_e} )$ be a multi-edge degree and let $\mathbf{x} := (x_1, \ldots, x_{n_e} )$ denote (vector) variables. We write $\mathbf{x^d}$ for $\prod_{i= 1}^{n_e} x_i^{d_i}$. Similarly, let $\mathbf{b} := (b_0, \ldots, b_{n_r} )$ be a received degree and let $\mathbf{r} := (r_0, \ldots, r_{n_r} )$ denote variables corresponding to received distributions. By $\mathbf{r^b}$ we mean $\prod_{i= 0}^{n_r} r_i^{b_i}$. In this paper we use $r_1$ for the transmission channel and $r_0$ for punctured bits (with no transmission channel). Typically, vectors $\mathbf{b}$ will have one entry set to $1$ and the rest set to $0$. Finally, the coefficients $\nu_{\mathbf{bd}}$ and $\mu_{\mathbf{d}}$, are non-negative real values corresponding to the fraction of variable nodes  of type $(\mathbf{bd})$ and the fraction of constraint nodes of type $\mathbf{d}$ in the graph.
\par For example, let $N$ be the length of the code-word, then for each constraint node degree type $\mathbf{d}$ the quantity $\mu_{\mathbf{d}}N$ is the number of constraint nodes of type $\mathbf{d}$ in the graph. Similarly, the quantity $\nu_{\mathbf{bd}}N$ is the number of variable nodes of type $(\mathbf{bd})$ in the graph. We store this information in a table to describe the structure of the graph. For instance a full description of a rate $0.02$ MET-LDPC code ensemble we designed with the following structure is presented in Table~\ref{table:rate002designed} and Fig. \ref{fig:rate002ensemble}.
\begin{align*}
    \nu(\mathbf{r,x}) =&~0.02\,r_1x_1^2x_2^{51} +0.02\,r_1x_1^3x_2^{60}+0.96\,r_1x_3~,\\
    \mu(\mathbf{x}) =& ~0.016\,x_1^{4} +~0.004\,x_1^{9}+~0.30\,x_2^{3}x_3^1+~0.66\,x_2^{2}x_3^1~.
\end{align*}
\begin{table}[!ht]
	\caption{Table presentation of the degree structure of a rate $0.02$ MET-LDPC code with $3$ edge types.}
	\begin{center}\label{table:rate002designed}	%\resizebox{0.91\hsize}{!}{
		\begin{tabular}{|c|c|c||c|c|}
			\hline
			$\nu_{\mathbf{bd}}$ & $\mathbf{b}$ & $\mathbf{d}$& $\mu_{\mathbf{d}}$& $\mathbf{d}$\\
			\hline
			$\begin{array}{c}
0.02  \\
0.02 \\
0.96 \end{array}$&[$0$ $1$]&$\begin{array}{ccc}
2&51&0  \\
3&60&0 \\
0&0&1 \end{array}$&$\begin{array}{c}
0.016  \\
0.004 \\
0.30  \\
0.66 \end{array}$&$\begin{array}{ccc}
4&0&0  \\
9&0&0 \\
0&3&1 \\
0&2&1\end{array}$\\
			\hline
			\hline
			\multicolumn{5}{c}{BIAWGN: $\sigma^*_\textsubscript{DE} = 5.94$}\\
			\hline
			\hline
		\end{tabular}
%}
	\end{center}
\end{table}
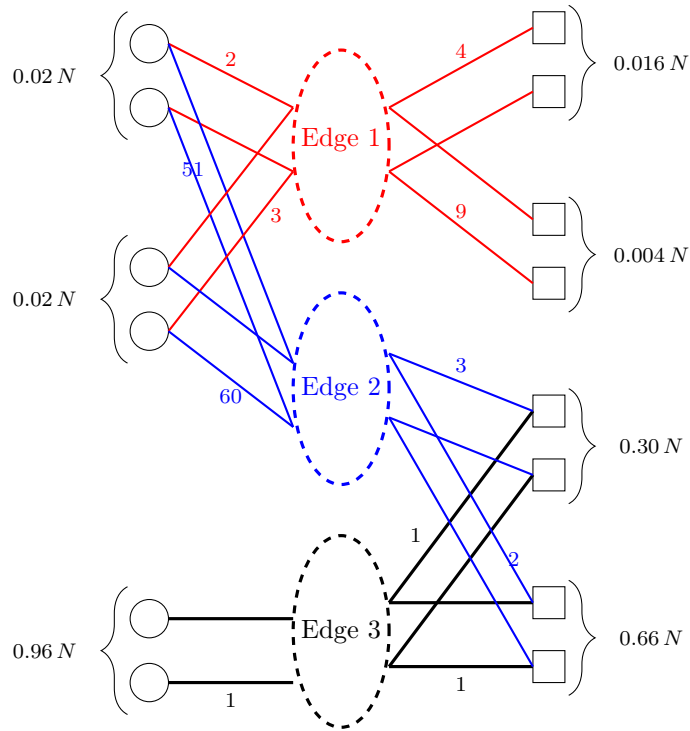
\begin{figure}[!htp]
	\begin{center}
%	\resizebox{0.9\hsize}{!}{
		\begin{tikzpicture}[scale=0.85]
		\def \Vn {3}
		\def \Cn {2}
		\def \Ne {3}
		\def \radius {0.3cm}
		\def \vertspace {3.8}

		\node at (0, 9.5) [red] {Edge 1};
		\draw [dashed, very thick, red] (0,9.4) ellipse (0.75cm and 1.5cm) ;
		\node at (0, 5.6) [blue] {Edge 2};
		\draw [dashed, very thick, blue] (0,5.6) ellipse (0.75cm and 1.5cm) ;
		\node at (0, 1.8) [black] {Edge 3};
		\draw [dashed, very thick, black] (0,1.8) ellipse (0.75cm and 1.5cm) ;
		\draw [decorate,decoration={brace,amplitude=10pt},xshift=-4pt,yshift=0pt]
		(-3.2,9.5) -- (-3.2,11.5) node [black,midway,xshift=-1.1cm] 
		{\footnotesize $0.02\,N$};
		\draw (-3,11) circle (\radius);
		\draw (-3,10) circle (\radius);
		\draw [decorate,decoration={brace,amplitude=10pt},xshift=-4pt,yshift=0pt]
		(-3.2,6) -- (-3.2,8) node [black,midway,xshift=-1.1cm] 
		{\footnotesize $0.02\,N$};
		\draw (-3,7.5) circle (\radius);
		\draw (-3,6.5) circle (\radius);
		\draw [decorate,decoration={brace,amplitude=10pt},xshift=-4pt,yshift=0pt]
		(-3.2,0.5) -- (-3.2,2.5) node [black,midway,xshift=-1.1cm] 
		{\footnotesize $0.96\,N$};
		\draw (-3,2) circle (\radius);
		\draw (-3,1) circle (\radius);
%		
%		
		% Check nodes
		\draw [decorate,decoration={brace, mirror, amplitude=10pt},xshift=-4pt,yshift=0pt]
		(3.7,9.8) -- (3.7,11.6) node [black,midway,xshift=1.1cm] 
		{\footnotesize $0.016\,N$};
		\draw (3,11) rectangle (3.5,11.5);
		\draw (3,10) rectangle (3.5,10.5);
		\draw [decorate,decoration={brace, mirror, amplitude=10pt},xshift=-4pt,yshift=0pt]
		(3.7,6.8) -- (3.7,8.6) node [black,midway,xshift=1.1cm] 
		{\footnotesize $0.004\,N$};
		\draw (3,8) rectangle (3.5,8.5);
		\draw (3,7) rectangle (3.5,7.5);
		\draw [decorate,decoration={brace, mirror, amplitude=10pt},xshift=-4pt,yshift=0pt]
		(3.7,3.8) -- (3.7,5.6) node [black,midway,xshift=1.1cm] 
		{\footnotesize $0.30\,N$};
		\draw (3,5) rectangle (3.5,5.5);
		\draw (3,4) rectangle (3.5,4.5);
		\draw [decorate,decoration={brace, mirror, amplitude=10pt},xshift=-4pt,yshift=0pt]
		(3.7,0.8) -- (3.7,2.6) node [black,midway,xshift=1.1cm] 
		{\footnotesize $0.66\,N$};
		\draw (3,2) rectangle (3.5,2.5);
		\draw (3,1) rectangle (3.5,1.5);
		%% Edges
		% from Vnode 28 to Edge 1
		\draw[solid,  thick,red] (-2.7,11) -- (-0.75, 10) node [red,above, midway,xshift=-0.0cm,yshift=0.0cm] {\footnotesize $2$};
		\draw[solid,  thick,red] (-2.7,10) -- (-0.75, 9);% node [red,midway,xshift=-0.6cm,yshift=0.6cm] {\footnotesize $3$};
		% from Vnode 28 to Edge 2
		\draw[solid,  thick,blue] (-2.7,11) -- (-0.75, 6); % node [blue,midway,xshift=-0.5cm,yshift=1.3cm] {\footnotesize $57$};
		\draw[solid,  thick,blue] (-2.7,10) -- (-0.75, 5) node [blue,midway,xshift=-0.5cm,yshift=1.3cm] {\footnotesize $51$};
		% from Cnode 17 to Edge 1
		\draw[solid,  thick,red] (0.75,10) -- (3, 11.25) node [red,above, midway,xshift=0.0cm,yshift=0.0cm] {\footnotesize $4$};
		\draw[solid,  thick,red] (0.75,9) -- (3, 10.25);% node [black,midway,xshift=0.6cm,yshift=0.6cm] {\footnotesize $3$};
		% from Cnode 15 to Edge 1
		\draw[solid,  thick, red] (0.75,9) -- (3, 7.25) node [red,midway, above, xshift=0.0cm,yshift=-0.0cm] {\footnotesize $9$};
		\draw[solid,  thick, red] (0.75,10) -- (3, 8.25); % node [black,midway,xshift=0.6cm,yshift=-0.6cm] {\footnotesize $15$};
		% from Vnode 36 to Edge 1
		\draw[solid,  thick,red] (-2.7,7.5) -- (-0.75, 10) ; % node [black,midway,xshift=0.6cm,yshift=0.7cm] {\footnotesize $2$};
		\draw[solid,  thick, red] (-2.7,6.5) -- (-0.75, 9) node [red, below, midway,xshift=0.6cm,yshift=0.7cm] {\footnotesize $3$};
		% from Vnode 36 to Edge 2
		\draw[solid, thick, blue] (-2.7,7.5) -- (-0.75, 6); % node [black,midway,xshift=0.0cm,yshift=0.0cm] {\footnotesize $57$};
		\draw[solid,  thick, blue] (-2.7,6.5) -- (-0.75, 5) node [blue, , below, midway,xshift=0.0cm,yshift=0.0cm] {\footnotesize $60$};
		% from Vnode 1536 to Edge 3
		\draw[solid, very thick, black] (-2.7,1) -- (-0.75, 1) node [black,midway, below] {\footnotesize $1$};
		\draw[solid, very thick, black] (-2.7,2) -- (-0.75, 2); % node [black,midway,below] {\footnotesize $1$};
		% from Cnode 576 to Edge 3
		\draw[solid, very thick, black] (3,4.25) -- (0.75, 1.25); % node [black,midway, below] {\footnotesize $1$};
		\draw[solid, very thick, black] (3,5.25) -- (0.75, 2.25) node [black,midway,above,xshift=-0.6cm,yshift=-0.6cm] {\footnotesize $1$};
		% from Cnode 960 to Edge 3
		\draw[solid, very thick, black] (3,1.25) -- (0.75, 1.25) node [black,midway, below] {\footnotesize $1$};
		\draw[solid, very thick, black] (3,2.25) -- (0.75, 2.25);% node [black,midway,below] {\footnotesize $1$};
		% from Cnode 960 to Edge 2
		\draw[solid,thick, blue] (3,4.25) -- (0.75, 5.15); % node [black,midway, above] {\footnotesize $2$};
		\draw[solid,thick, blue] (3,5.25) -- (0.75, 6.15) node [blue,midway,above] {\footnotesize $3$};
        % from Cnode 576 to Edge 2
		\draw[solid,thick, blue] (3,1.25) -- (0.75, 5.15); % node [blue,midway,xshift=0.7cm,yshift=-1.3cm] {\footnotesize $3$};
		\draw[solid,thick, blue] (3,2.25) -- (0.75, 6.15) node [blue,midway, above, xshift=0.7cm,yshift=-1.3cm] {\footnotesize $2$};
		\end{tikzpicture}
%		}
	\end{center}\caption{Graphical representation of a three-edge type-LDPC code presented in Table~\ref{table:rate002designed}, where $\bigcirc$ represents the variable nodes and $\square$ represents the check nodes. The number of nodes for different edge-types are shown as fractions of the code length $N$, where $N$ is the number of transmitted code-word bits.}\label{fig:rate002ensemble}
\end{figure}
\par The edge perspective degree distribution can be described as a vector of multi-variable polynomials, for variable nodes and check nodes, respectively,
\begin{align}\label{eq:dd4MELDPC-Edgeperspective}
\nonumber \mathbf{\lambda(r,x)} &= \left( \dfrac{\nu_{x_1}(\mathbf{r,x})}{\nu_{x_1}(\mathbbm{1,1})},\, \dfrac{\nu_{x_2}(\mathbf{r,x})}{\nu_{x_2}(\mathbbm{1,1})}, \ldots, \dfrac{\nu_{x_{n_e}}(\mathbf{r,x})}{\nu_{x_{n_e}}(\mathbbm{1,1})} \right)\\
\mathbf{\rho(x)} &= \left( \dfrac{\mu_{x_1}(\mathbf{x})}{\mu_{x_1}(\mathbbm{1})},\, \dfrac{\mu_{x_2}(\mathbf{x})}{\mu_{x_2}(\mathbbm{1})},\, \ldots, \dfrac{\mu_{x_{n_e}}(\mathbf{x})}{\mu_{x_{n_e}}(\mathbbm{1})} \right)\ ,
\end{align}
where 
\begin{eqnarray}
\nonumber \nu_{x_i}(\mathbf{r,x})  = \frac{\partial }{\partial x_i} \nu(\mathbf{r,x}),~~~~ \mu_{x_i}(\mathbf{x})  = \frac{\partial }{\partial x_i} \mu(\mathbf{x}),
\end{eqnarray}
and $\mathbbm{1}$ denotes a vector of all 1’s where the length is being determined by context.
The coefficients of $\nu$ and $\mu$ are constrained to ensure that the number of sockets of each type is the same on both sides (variable and check) of the graph. This gives rise to $n_e$ linear conditions on the coefficients of $\nu$ and $\mu$,
\begin{equation*}
\nu_{x_i}(\mathbbm{1,1}) = \mu_{x_i}(\mathbbm{1}), \mbox{   } i = 1, \ldots, n_e\ .
\end{equation*}
Finally, the nominal code rate for non-punctured code is given by
\begin{equation*}
R^{ch} = \nu(\mathbbm{1,1}) - \mu(\mathbbm{1})\ .
\end{equation*}
%%%%%%%%%%%%%%%%%%%%%%%%%%%%%%%%%%%%%%%%%%%%%% 
%%%%%%%%%%%%%%%%%%%%%%%%%%%%%%%%%%%%%%%%%%%%%% SEC III
%%%%%%%%%%%%%%%%%%%%%%%%%%%%%%%%%%%%%%%%%%%%%% 
\section{Generalized Extrinsic Information Transfer Chart}\label{sec:G-EXIT}
\subsection{Belief Propagation and asymptotic analysis tools}\label{sec_5}
Density evolution (DE) is the main tool for analyzing the average asymptotic behavior of the belief propagation (BP) decoders for MET-LDPC code ensembles with infinite block length and infinite number of iterations. The DE analysis is in general simplified by the all-one code word assumption, the channel symmetry and by going to the log-likelihood ratio (LLR) domain~\cite{7544625Johnson,Saeedi2010,Urbanke2005}. Let us denote by  $\mathbf{P} = (P_1, ..., P_{n_e} )$, vectors of symmetric densities where $P_i$ is the density of messages carried on edge type $i$. Also assume that $\mathbf{P}^l$ denote the vector of messages passed from variable nodes to check nodes in iteration $l$ assuming that $\mathbf{P}^0 = \mathbf{\Delta_0}$. By $\mathbf{\Delta_0}$ we mean a vector of densities where each density is $\delta_0$, which is equivalent to erasure with probability $1$. Similarly, let $\mathbf{R}$ be the received distributions. Then the following recursion represents the density evolution for MET-LDPC codes:
\begin{align}\label{eq:DE4METLDPC}
\mathbf{P}^{l+1} &= \mathbf{\lambda}(\mathbf{R},\mathbf{\rho}(\mathbf{P}^{l}))\ ,\\
\mathbf{Q}^{l+1} &= \mathbf{\rho}(\mathbf{P}^{l})\ ,
\end{align}
where, $\mathbf{\rho}(\mathbf{x})$ and $\mathbf{\lambda(\mathbf{r},\mathbf{x})}$ are presented in Eq.~\eqref{eq:dd4MELDPC-Edgeperspective}. Detailed calculation of the density evolution for MET-LDPC codes can be found in Section II-B of~\cite{7544625Johnson}. 
\subsection{Generalized-EXIT function, G-EXIT curve and Dual G-EXIT curve}
\par The original idea behind the G-EXIT chart method is to demonstrate the decoding process using a suitable one-dimensional representation of the densities~\cite{measson2006conservation}. The G-EXIT chart is visualized on the basis of two G-EXIT curves that represent the action of the different types of nodes. Considering the fact that for MET-LDPC codes, the DE tracks $n_e$  message densities as presented in Eqs.~\eqref{eq:dd4MELDPC-Edgeperspective} and \eqref{eq:DE4METLDPC}, the G-EXIT chart for MET-LDPC codes is also expanded to a vector of $n_e$ components. This makes the G-EXIT analysis tools unpractical when the number of edges are more than three $(n_e > 3)$. Intuitively we present again a one-dimensional G-EXIT chart by exploiting appropriate convolution in variable nodes and constraint nodes before applying the G-EXIT projection to the densities. 
\par Based on the results of~\cite{Measson2009GAT}, given two families of $L$-densities $\{c_{\epsilon_i}\}$ and $\{a_{\epsilon}\}$ parameterized by $\epsilon$, the G-EXIT function can be represented as
% \begin{small}
\begin{equation}\label{func: normalized G-EXIT no channel}
%\resizebox{0.85\hsize}{!}{$
G(c_{\epsilon_i}, \,
a_\epsilon)=\,\frac{\int_{z}\int_{w}{a_\epsilon(z)\frac{\partial
			c_{\epsilon_i}(\omega)}{\partial
			\epsilon}}\log_2\left(1+e^{-z-\omega}\right) d\omega
	dz}{\int_{w}\frac{\partial c_{\epsilon_i}(\omega)}{\partial
		\epsilon}\log_2\left(1+e^{-\omega}\right) d\omega}\ ,
%$}
\end{equation}
and the G-EXIT kernel is defined as
\begin{equation}\label{func: normalized kernel}
%\resizebox{0.8\hsize}{!}{$
l^{c_{\epsilon_i}}(z)=\frac{\int_\omega \frac{\partial
		c_{\epsilon_i}}{\partial \epsilon} \log_2{\left(1+
		e^{-z-w}\right)} d\omega}{\int_\omega \frac{\partial
		c_{\epsilon_i}}{\partial \epsilon} \log_2{\left(1+
		e^{-w}\right)}d\omega}\ .
%$}
\end{equation}
% \end{small}
Consequently, the G-EXIT curve is given in parametric form by $\{H(c_{\epsilon_i}), \, G(c_{\epsilon_i}, \, a_\epsilon)\}$, where $$H(c_{\epsilon_i})= \int_{-\infty}^{\infty}c_{\epsilon_i}(\omega)\log_2(1+e^{-\omega})d\omega\ .$$
\par
According to Eq.~\eqref{eq:DE4METLDPC}, the DE provides two vectors with $n_e$-components of densities for the variable nodes and check nodes, respectively. In order to plot the one-dimensional G-EXIT chart, these $n_e$ densities corresponding to each edge type will be combined to a single family of densities based on Eq.~\eqref{eq:dd4MET-LDPC}. Thus for the MET-LDPC codes the combination of densities for variable nodes and check nodes with $n_e$ edges are
\begin{eqnarray}\label{eq:intermeditedensities}
\label{eq:intermeditedensities1} c_{\epsilon_i} &=& \sum \nu_{\mathbf{bd}} \mathbf{R^ {\otimes b}} \otimes [\mathbf{Q}^{l+1}]^{\otimes \mathbf{d}}\ ,\\ %\bigotimes_{j=1}^{n_e}{c^j_{\epsilon_i}}\\
\label{eq:intermeditedensities2} a_\epsilon &=&  \sum \mu_{\mathbf{d}}[\mathbf{P}^{l}]^{\boxtimes \mathbf{d}}\ , %{\boxtimes}_{j=1}^{n_e}{a^j_\epsilon}\ ,
\end{eqnarray}
where $\mathbf{Q}^{\otimes \mathbf{d}}$ denotes $\bigotimes_{i=1}^{n_e} Q_i^{\otimes d_i}$, similarly $\mathbf{R^b}$ denotes $\bigotimes_{j=0}^{n_r} R_j^{\otimes b_j}$, and $\otimes$ denotes convolution in variable nodes. In a similar way $\mathbf{P^ {\boxtimes d}}$ denotes ${\boxtimes}_{i=1}^{n_e}{P_i^{\boxtimes d_i}}$, and $\boxtimes$ denote the convolution of check nodes. According to~\cite{Measson2009GAT} the dual G-EXIT curve is defined in parametric form as $\{G(a_\epsilon, \, c_{\epsilon_i}), \, H(a_{\epsilon})\}$, where we can use \eqref{func: normalized G-EXIT no channel} and swap the arguments to calculate that.
It is proven that for a binary linear code and transmission over \textit{Binary Memoryless Symmetric} (BMS) channels that the G-EXIT and the dual G-EXIT curve have equal area~\cite{Measson2009GAT}.
%%%%%%%%%%%%%%%%%%%%%%%%%%%%%%%%%%%%
%%%%%%%%%%%%%%%%%%%%%%%%%%%%%%%%%%%%
%%%%%%%%%%%%%%%%%%%%%%%%%%%%%%%%%%%%
\section{G-EXIT charts for code design}\label{sec:Code Design}
\subsection{Examples of G-EXIT charts for MET-LDPC codes}
In this section we present some examples of MET-LDPC codes and try to find the threshold of the codes using the G-EXIT chart method. We start with the rate $0.02$ MET-LDPC code in Table~\ref{table:rate002designed}. The Shannon limit for rate $0.02$ is equal to ${E_b}/{N_0} = -1.53$ dB ($\sigma^*_\textsubscript{Sh} =  5.96$) and our proposed code has a threshold equal to $-1.5$ dB ($\sigma^*_\textsubscript{DE} = 5.94$) which is just $0.03$ dB away from capacity. With $E_b$ being the energy per bit and $N_0$ being the energy of the noise, the relation between $E_b/N_0$, the SNR and $\sigma$ for an AWGN channel with binary transmission is (linear scale) 
%-15.51   -15.48%
\begin{align}
\nonumber \mbox{SNR } &= 2 R^{ch}\dfrac{E_b}{N_0}\ ,\\
\nonumber \sigma &= \dfrac{1}{\sqrt{\mbox{SNR}}}\ .
\end{align} 
\par
\begin{figure}[!ht]
	\centering
	\subfigure[First edge type.\label{fig:a}]{
		\includegraphics[width=0.42\textwidth]{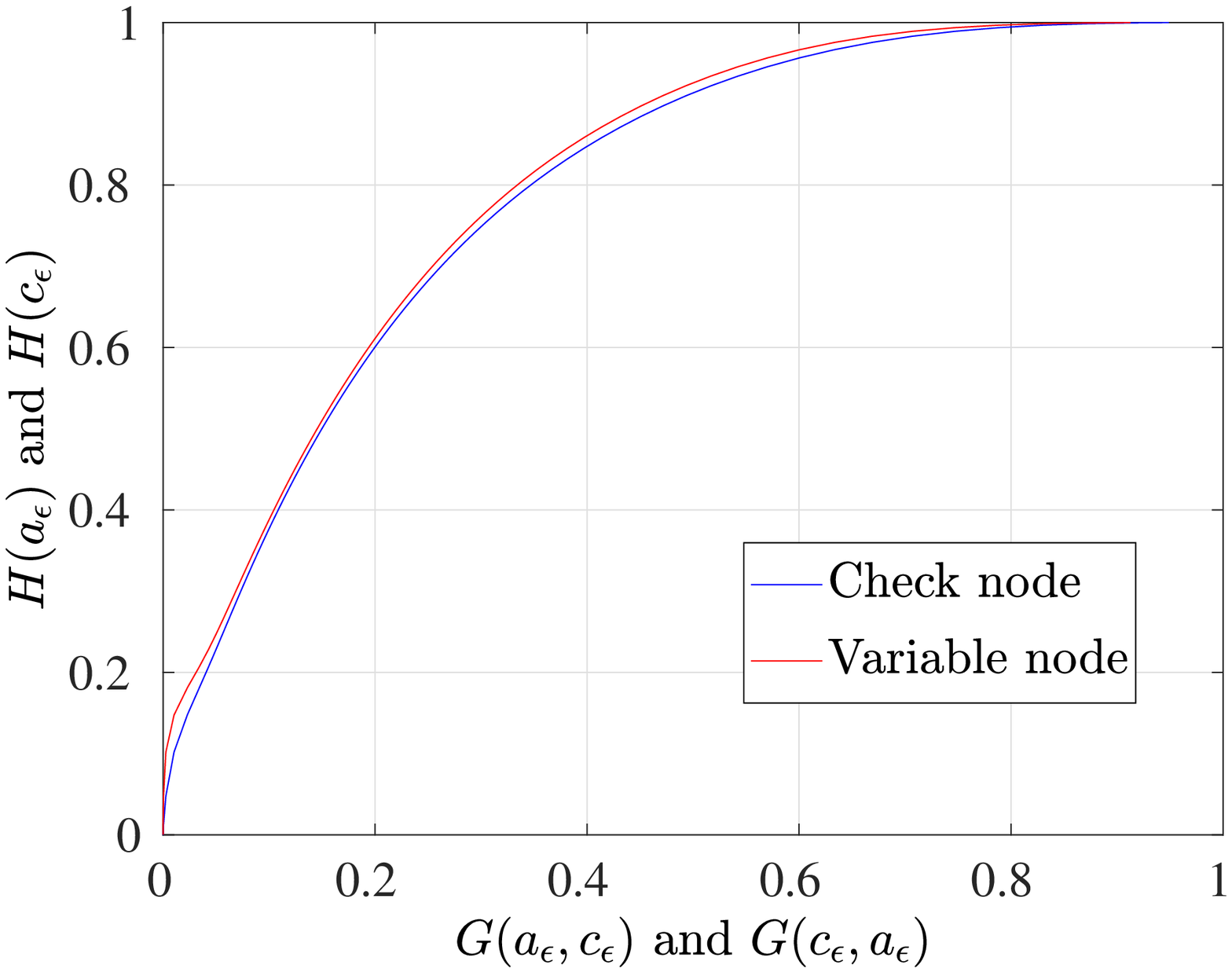}
	}
	\subfigure[Second edge type.\label{fig:b}]{
		\includegraphics[width=0.43\textwidth]{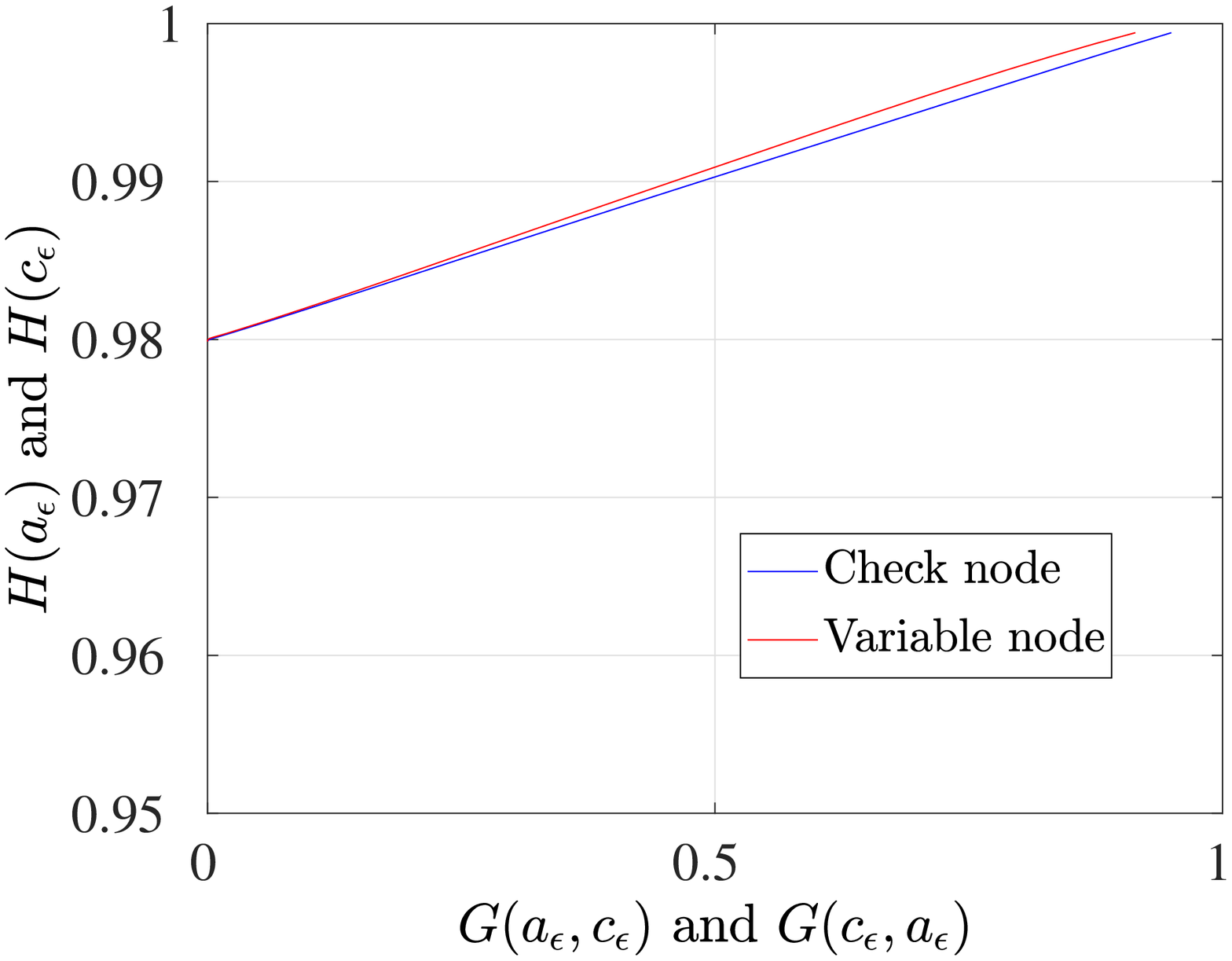}
	}\hfill
	\subfigure[Third edge type.\label{fig:c}]{
		\begin{tikzpicture}[spy using outlines={circle,gray,magnification=9.,size=2.5cm, connect spies}]
		\node {\includegraphics[width=0.44\textwidth]{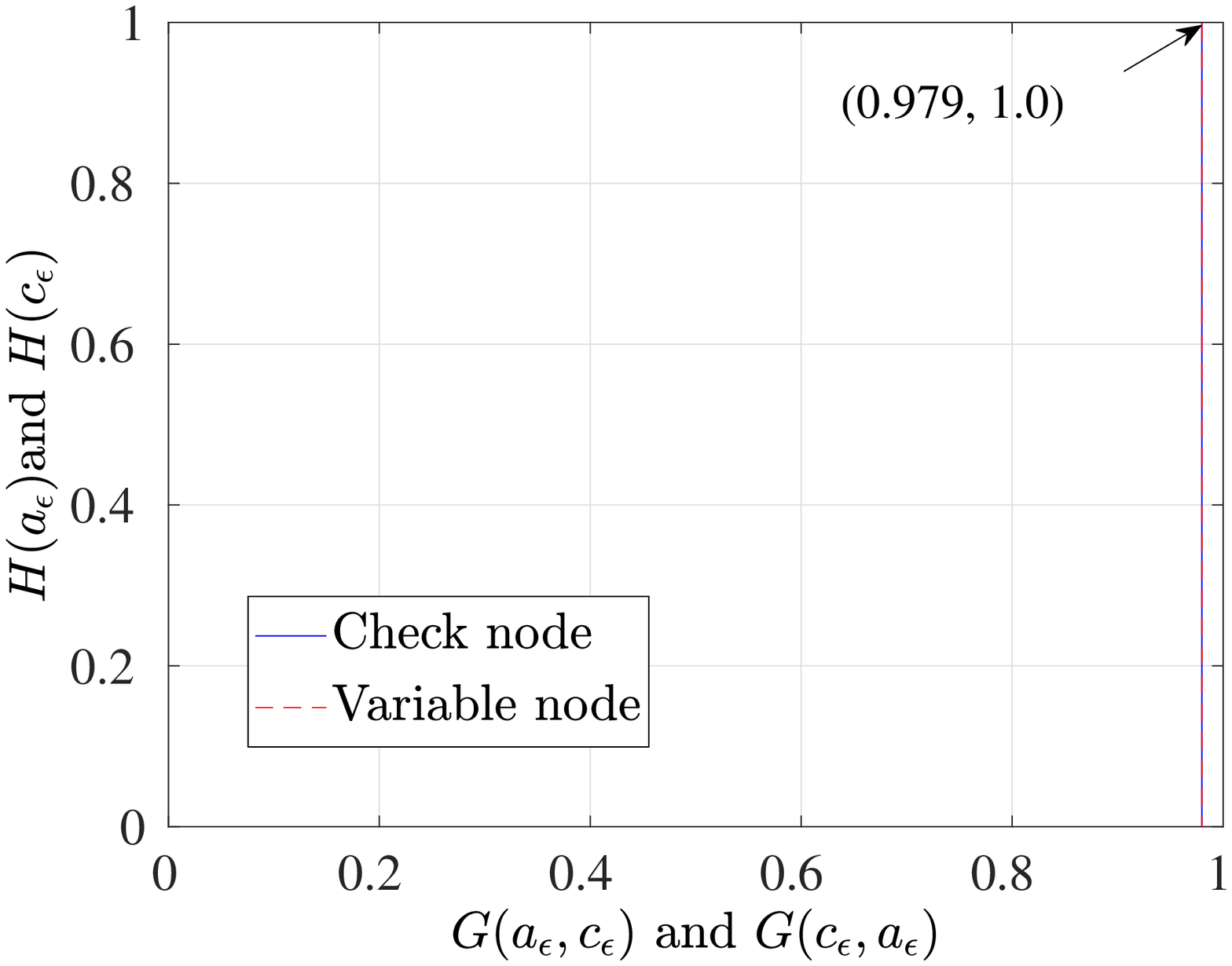}};
		\spy on (2.77,-0.15) in node [left] at (1.1,0.6); %on (3.9,-0.2) in node [left] at (.5,-0.2);
		\end{tikzpicture}
	}\hfill
	\caption{The G-EXIT charts for separate edges for the rate $0.02$ MET-LDPC code detailed in Table~\ref{table:rate002designed}. The SNR of this code is $-15.48$ dB ($\sigma^*_\textsubscript{DE} = 5.94$). When the code converges at a specific threshold value we are able to plot the G-EXIT charts separately by applying the G-EXIT operators for densities at each edge.}\label{fig:G-EXITrate002Separate}
\end{figure}

 Also, using  Eq.~\eqref{eq:dd4MELDPC-Edgeperspective} the density evolution vector of multi-variable polynomials can be written as
\begin{eqnarray}\label{eq:dd4MELDPC-rate2-100}
\nonumber \mathbf{\lambda(r,x)} &=& \biggl[0.6\,r_1x_1^2x_2^{60}+0.4\,r_1x_1x_2^{51},~0.54505\,r_1x_1^3x_2^{59}+0.4595\,r_1x_1^2x_2^{50},~r_1 \biggl]\ ,\\
\nonumber \mathbf{\rho(x)} &=& \biggl[0.64\,x_1^{3}+0.36\,x_1^8,\,~0.4054\,x_2^2x_3+0.5946\,x_2x_3,~0.3125\,x_2^3+0.6875\,x_2^2\biggl]\ ,
\end{eqnarray}
where by replacing the vectors of variables of $\mathbf{r}$ and $\mathbf{x}$ with vectors of densities $\mathbf{R}$, $\mathbf{P}$ and $\mathbf{Q}$ for channel, check nodes and variable nodes, respectively, we have the MET-DE. Figure~\ref{fig:G-EXITrate002Separate} shows the convergence behavior of each edge for the above mentioned code. 

It is noteworthy to mention that there is a single edge type with degree one variable nodes for this MET-LDPC code (c.f. Table~\ref{table:rate002designed}). These nodes apply a fixed channel density at each iteration of the DE. The corresponding G-EXIT curve for this edge is plotted in Fig.~\ref{fig:c}, which is constructed from two completely matching vertical lines at a specific $x$-value which denotes the entropy of the channel $H(\sigma^*_\textsubscript{DE}) = 0.9799$. 

\begin{figure}[!ht]
	\begin{center}
	\includegraphics[width=0.6\textwidth,keepaspectratio]{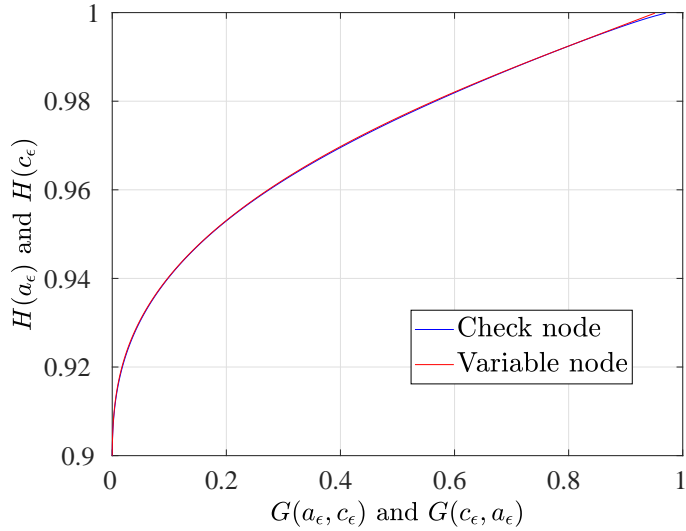}
	\end{center}\caption{The G-EXIT chart for rate the $0.02$ MET-LDPC code detailed in Table~\ref{table:rate002designed}. The SNR for this code is $-15.48$ dB which is equivalent to $\sigma^*_\textsubscript{DE} =  5.94$. 
	}\label{fig:G-EXITrate002}
\end{figure}

Finally, to see the convergence of a MET-LDPC code in a single plot, we used the overall combination of the edges with appropriate combination in check nodes and variable nodes according to Eqs.~\eqref{eq:intermeditedensities1}-\eqref{eq:intermeditedensities2}. The results are presented in Fig.~\ref{fig:G-EXITrate002}.

\par
\begin{figure}[!ht]
	\begin{center}
	\includegraphics[width=0.6\textwidth,keepaspectratio]{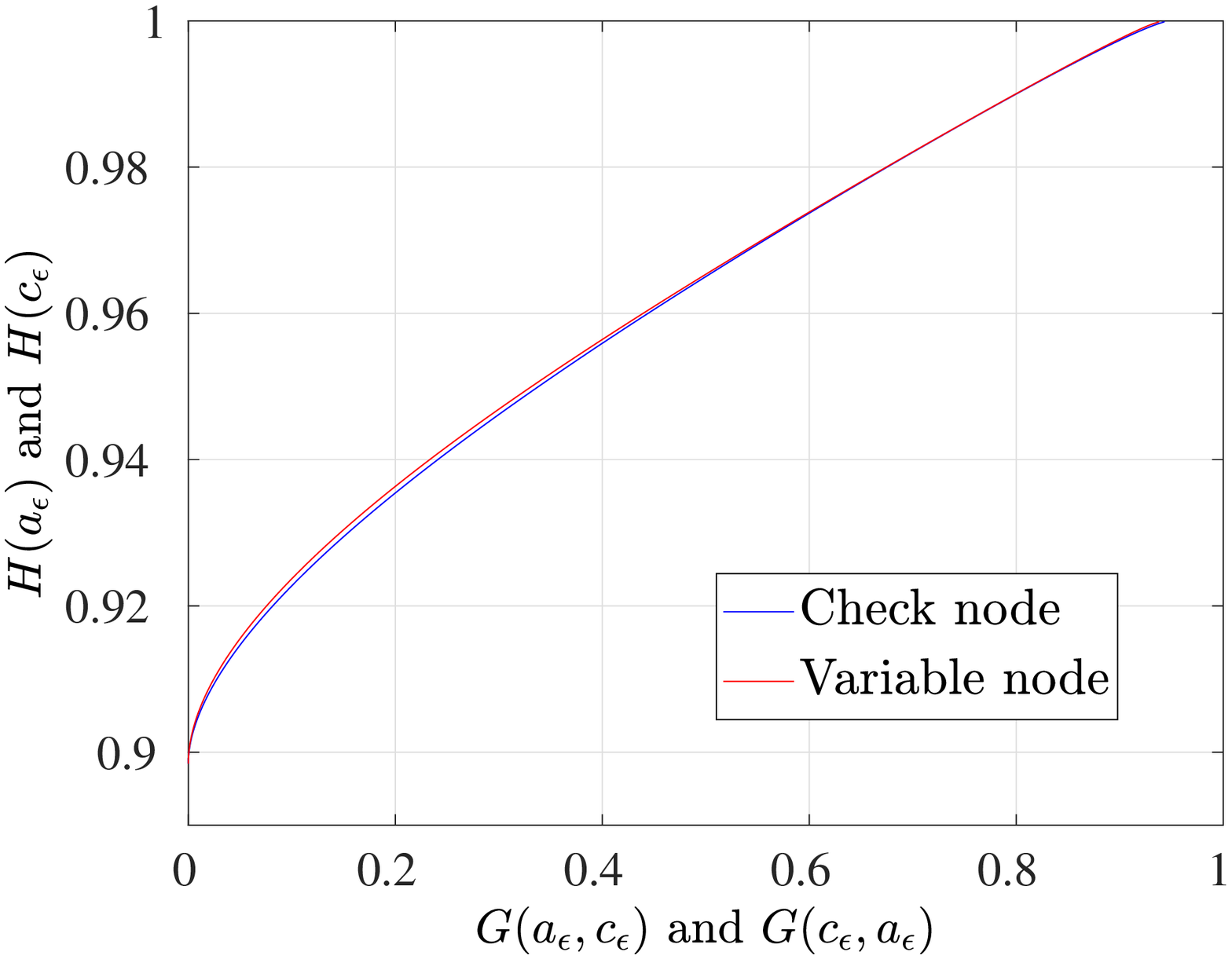}
	\end{center}\caption{The G-EXIT chart for rate the $0.05$ MET-LDPC code detailed in Table~\ref{table:rate005urbankemodified}. The SNR for this code is $-11.32$ dB which is equivalent to $\sigma^*_\textsubscript{DE} =  3.68$. %The red line is for the variable node and the blue line stands for the check node.
	}\label{fig:G-EXITrate005}
\end{figure}

 As a second example, Table~\ref{table:rate005urbankemodified} shows the degree structure of a rate $0.05$ MET-LDPC code. The threshold of this code using DE on BIAWGN channel is equal to $E_b/N_0 = -1.32$ dB ($\sigma^*_\textsubscript{DE} = 3.68$). The Shannon limit for a code of rate $0.05$ is equal to $-1.44$ dB ($\sigma^*_\textsubscript{Sh} = 3.73$) and this code is just $0.12$ dB away from capacity. The G-EXIT chart for this code is plotted in Fig. \ref{fig:G-EXITrate005}.
\begin{table}[!ht]
	\caption{Table presentation of the degree structure of a rate $0.05$ MET-LDPC code.}
	\begin{center}\label{table:rate005urbankemodified}
	%\resizebox{0.91\hsize}{!}{
		\begin{tabular}{|c|c|c||c|c|}
			\hline
			$\nu_{\mathbf{bd}}$ & $\mathbf{b}$ & $\mathbf{d}$& $\mu_{\mathbf{d}}$& $\mathbf{d}$\\
			\hline
			$\begin{array}{c}
0.054  \\
0.046 \\
0.90 \end{array}$&[$0$ $1$]&$\begin{array}{ccc}
2&22&0  \\
3&22&0 \\
0&0&1 \end{array}$&$\begin{array}{c}
0.026  \\
0.024 \\
0.40  \\
0.50 \end{array}$&$\begin{array}{ccc}
3&0&0  \\
7&0&0 \\
0&3&1 \\
0&2&1\end{array}$\\
			\hline
			\hline
			\multicolumn{5}{c}{BIAWGN: $\sigma^*_\textsubscript{DE} = 3.68$}\\
			\hline
			\hline
		\end{tabular}
%		}
	\end{center}
\end{table}

\begin{figure}[!ht]
	\begin{center}
	\includegraphics[width=0.6\textwidth,keepaspectratio]{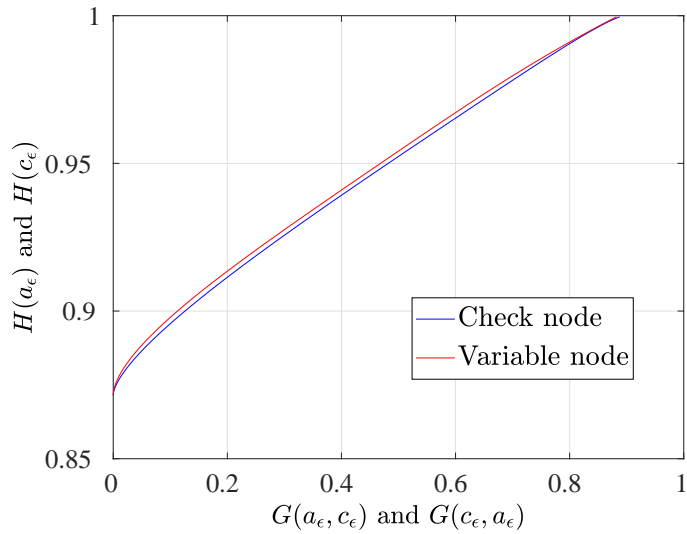}
\end{center}\caption{The G-EXIT chart for rate the $0.1$ MET-LDPC code detailed in Table~\ref{table:rate01urbanke}. The SNR for this code is $-8.21$ dB which is equivalent to $\sigma^*_\textsubscript{DE} =  2.57$. The corresponding curves are plotted at SNR $-8.07$ dB.}\label{fig:G-EXITrate01} %The red line is for the variable node and the blue line stands for the check node.
\end{figure}

As a third example, Fig.~\ref{fig:G-EXITrate01} demonstrates the G-EXIT chart for a rate $0.1$ MET-LDPC code. The node perspective degree structure of this code is presented in Table~\ref{table:rate01urbanke}.

\begin{table}[!ht]
	\caption{Table presentation of the degree structure of a rate $0.1$ MET-LDPC code.}
	\begin{center}\label{table:rate01urbanke}
	%\resizebox{0.91\hsize}{!}{
\begin{tabular}{|c|c|c||c|c|}
			\hline
			$\nu_{\mathbf{bd}}$ & $\mathbf{b}$ & $\mathbf{d}$& $\mu_{\mathbf{d}}$& $\mathbf{d}$\\
			\hline
			$\begin{array}{c}
0.075  \\
0.050 \\
0.875 \end{array}$&[$0$ $1$]&$\begin{array}{ccc}
2&23&0  \\
3&18&0 \\
0&0&1 \end{array}$&$\begin{array}{c}
0.025  \\
0.875  \end{array}$&$\begin{array}{ccc}
12&0&0  \\
0&3&1 \end{array}$\\
			\hline
			\hline
			\multicolumn{5}{c}{BIAWGN: $\sigma^*_\textsubscript{DE} = 2.57$}\\
			\hline
			\hline
		\end{tabular}
		%}
	\end{center}
\end{table}
The code has $n_e = 3$ edge type and the threshold of this code in a BIAWGN channel using DE is equal to $E_b/N_0 = -1.22$ dB ($\sigma^*_\textsubscript{DE} = 2.57$). The Shannon limit is equal to $-1.28$ dB ($\sigma^*_\textsubscript{Sh} = 2.6$) and this code is just $0.06$ dB away from capacity. The corresponding G-EXIT curve for this code is plotted in Fig. \ref{fig:G-EXITrate01}.
\par
\begin{figure}[!ht]
	\begin{center}
	\includegraphics[width=0.6\textwidth,keepaspectratio]{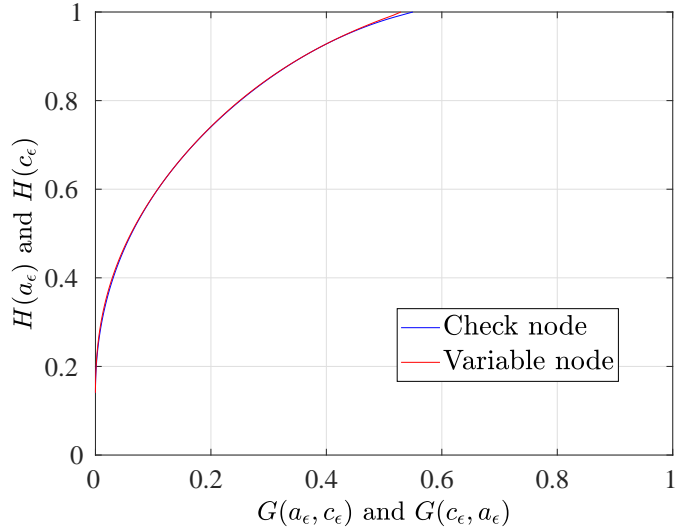}
	\end{center}\caption{The G-EXIT chart for rate the $0.5$ MET-LDPC code detailed in Table~\ref{table:rate05urbanke}. The SNR for this code is $ 0.305$ dB which is equivalent to $\sigma^*_\textsubscript{DE} =  0.9655$. 
	}\label{fig:G-EXITrate05}
\end{figure}

 As fourth example, Fig.~\ref{fig:G-EXITrate05} shows the G-EXIT chart for a rate $0.5$ MET-LDPC code presented in Table~\ref{table:rate05urbanke}. The code has three edge types and it also has punctured nodes. This code was first published in~\cite{richardson2002multi} and has a threshold equal to $E_b/N_0 = 0.305$ dB ($\sigma^*_\textsubscript{DE} = 0.9655$). The Shannon limit for a code of rate $0.5$ on BIAWGN channel is equal to $\sigma^*_\textsubscript{Sh} = 0.9787$.
\begin{table}[!ht]
	\caption{Table presentation of the degree structure of a rate $0.5$ MET-LDPC code~\cite{richardson2002multi}.}
	\begin{center}\label{table:rate05urbanke}
	%\resizebox{0.91\hsize}{!}{
\begin{tabular}{|c|c|c||c|c|}
			\hline
			$\nu_{\mathbf{bd}}$ & $\mathbf{b}$ & $\mathbf{d}$& $\mu_{\mathbf{d}}$& $\mathbf{d}$\\
			\hline
			$\begin{array}{c}
0.2  \\
0.5 \\
0.3\\
0.2 \end{array}$&$\begin{array}{cc}
1& 0  \\
0& 1 \\
0& 1\\
0 &1 \end{array}$&$\begin{array}{cccc}
0&3&3&0  \\
2&0&0&0 \\
3&0&0&0 \\
0&0&0&1\end{array}$&$\begin{array}{c}
0.1  \\
0.4 \\
0.2 \end{array}$&$\begin{array}{cccc}
3&2&0&0 \\
4&1&0&0\\
0&0&3&1 \end{array}$\\
			\hline
			\hline
			\multicolumn{5}{c}{BIAWGN: $\sigma^*_\textsubscript{DE} = 0.9655$}\\
			\hline
			\hline
		\end{tabular}
%}
	\end{center}
\end{table}
\subsection{Convergence behavior using the G-EXIT chart}
\par
\begin{figure}[!ht]
	\centering
	\subfigure[G-EXIT chart at $E_b/N_0 = 0.005$ dB.\label{fig:Conv_c}]{
	\includegraphics[width=0.47\textwidth]{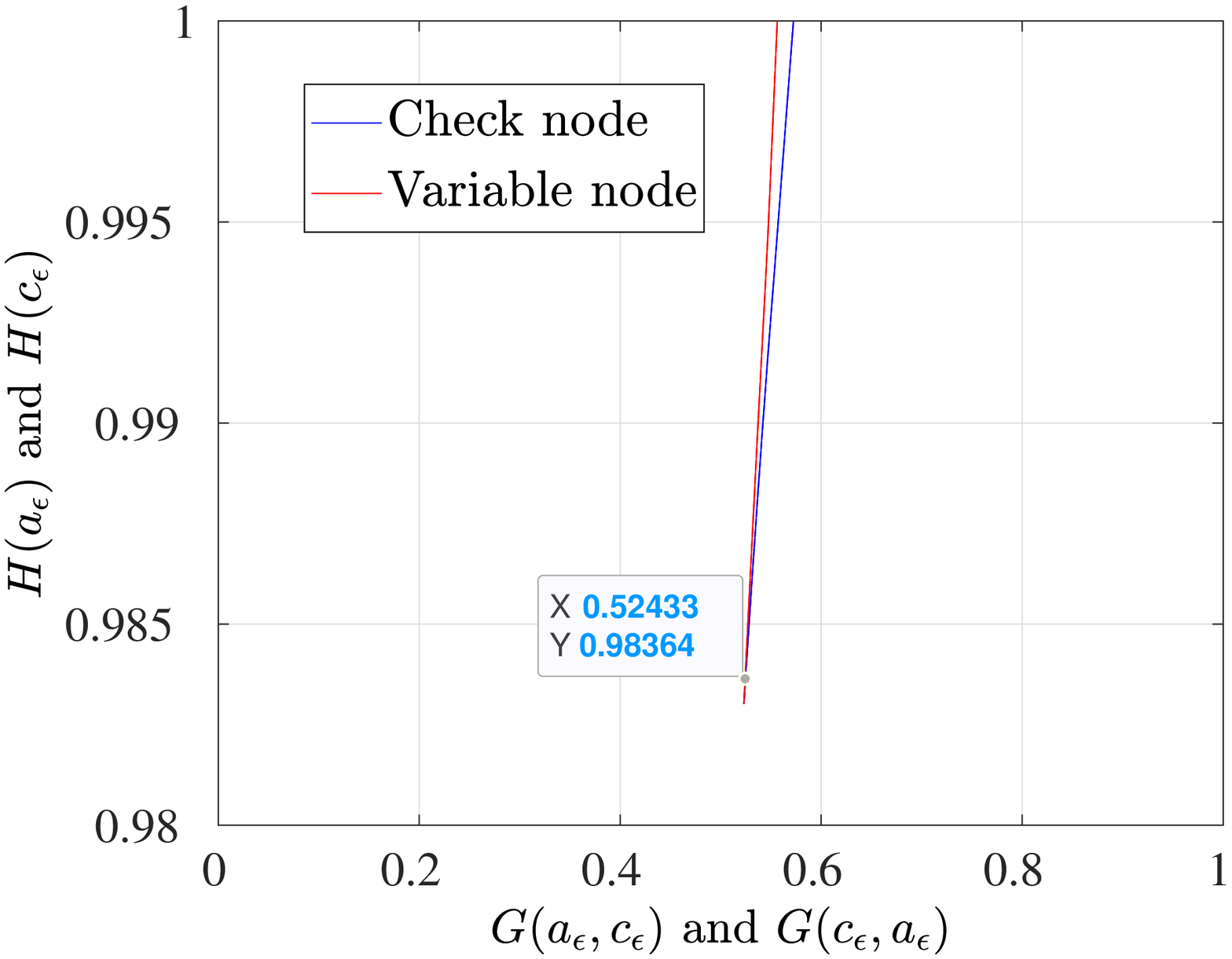}
	}
	\subfigure[G-EXIT chart at $E_b/N_0 = 0.63$ dB.\label{fig:Conv_a}]{
		\includegraphics[width=0.47\textwidth]{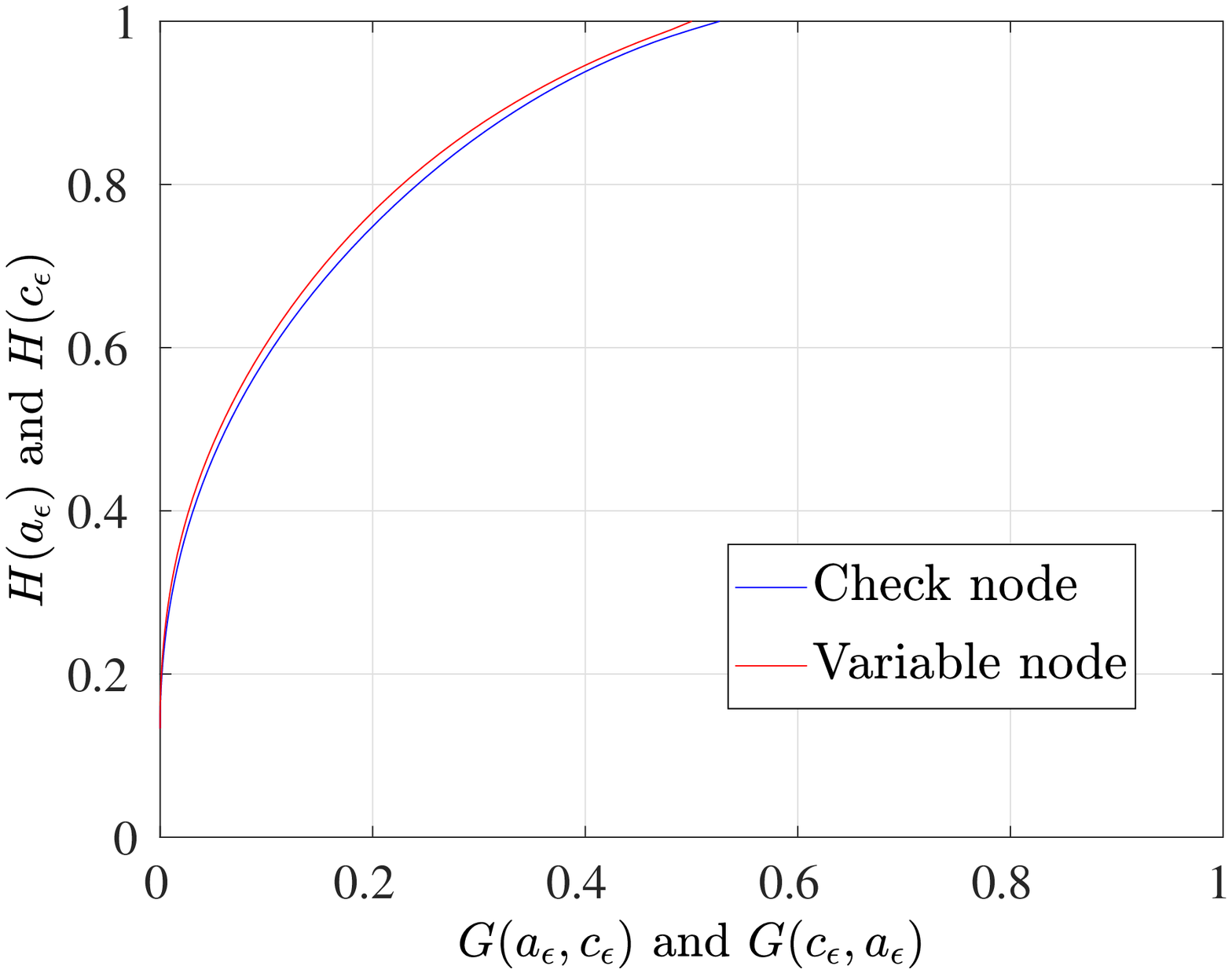}
	}\hfill
	\caption{The convergence behavior of the G-EXIT charts for rate $0.5$ MET-LDPC code. Figures \ref{fig:Conv_c} and \ref{fig:Conv_a} show the convergence behavior of the code for a noise variance of $\sigma = 0.9994$ and $\sigma = 0.9327$, respectively. For the former the code does not converge which can be seen by the crossing lines, while for the latter the code converges.}
	\label{fig:G-EXITConvergence}
\end{figure}

 Now we use the graphical presentation to demonstrate the convergence behavior of the code structure which we will exemplify for the rate $0.5$ MET-LDPC code (c.f.~\ref{table:rate05urbanke}). As depicted in Fig.~\ref{fig:G-EXITrate05}, the two curves are matched to each other and the threshold of this code is $0.305$ dB. In Fig.~\ref{fig:G-EXITConvergence} the G-EXIT charts are plotted for this code for two different $E_b/N_0 \in \{0.605, 0.005\}$, which are ${E_b/N_0}^{DE}\pm 0.3$. It is possible to translate convergence behavior of the code by monitoring the status of the G-EXIT curves. 
\par  For $E_b/N_0$ smaller than the threshold the code is not able to correct errors. In this case the two curves cross each other in the G-EXIT chart, see Fig.~\ref{fig:Conv_c}. For $E_b/N_0 = 0.63$\,dB, a value larger than the threshold, the corresponding G-EXIT chart is plotted in Fig.~\ref{fig:Conv_a}. The extra gap between the curves in comparison with Fig.~\ref{fig:G-EXITrate05} shows that the corresponding MET-LDPC code is still able to correct the errors even at lower $E_b/N_0$.
%%%%%%%%%%%%%%%%%%%%%%%%%%%%%%%%%%%%%%%%%%%%%%%%%%%%
\subsection{Gaussian assumption and complexity reduction}
For SET-LDPC codes for the BI-AWGN channel the well-known one dimensional Gaussian approximation can be used to determine the convergence threshold~\cite{design_BIAWGN_2, Ardakani2004, Divsalar2009_Capacityapproaching, Shokrollahi2000DesignBEC, chung2000construction, Urbanke2001capacity, Johnson2014_Optimization}. In case the check node degrees are small and the variable degrees are large enough, the PDF of both the variable and the check nodes can be approximated by a Gaussian distribution for all input and intermediate densities \cite{7544625Johnson}. The Gaussian PDF is thereby determined by its mean. It is thus enough to trace only a single parameter during the BP decoding algorithm.

For MET-LDPC this Gaussian approximation is not valid~\cite{7544625Johnson}. In this part we introduce a new analysis tool for MET-LDPC codes on AWGN channels which is significantly more accurate than the conventional Gaussian approximation. In our proposed method we assume a Gaussian distribution only for messages from variable nodes to check nodes. In comparison to other existing methods which assume Gaussian approximation for both check nodes and variable nodes~\cite{7544625Johnson}, our method calculates the check node PDFs based on check node operations. To show the accuracy of this method we combined the G-EXIT operator to our approximation method and found the threshold and convergence behavior of the codes. Simulation results show that our proposed method provides an accurate estimate of the convergence behavior and the threshold of the code. 
\par
\begin{figure}[!ht]
	\centering
	\subfigure[Intermediate densities at the variable node outputs.\label{fig:VarDE_a}]{
		\includegraphics[width=0.48\textwidth,keepaspectratio]{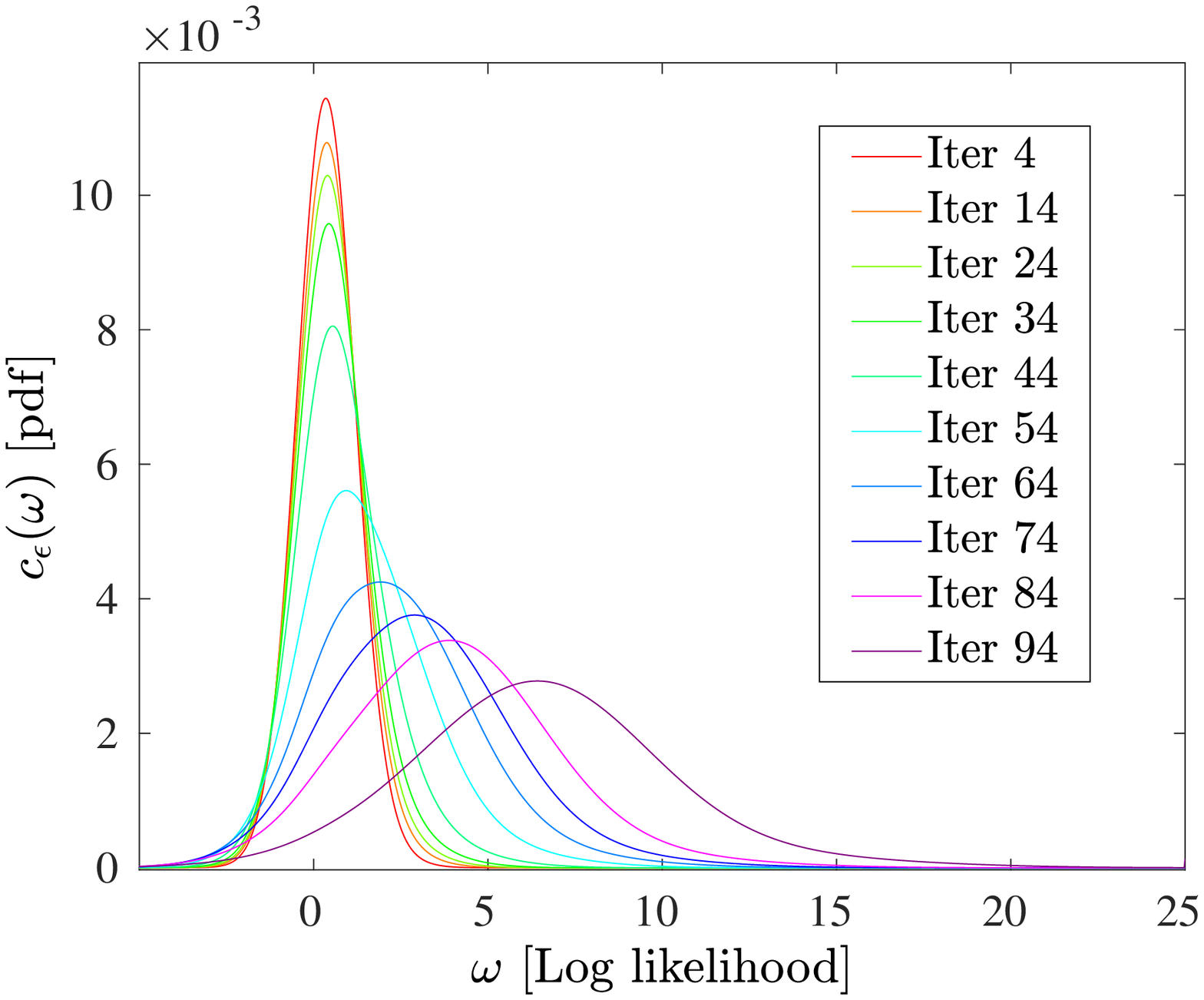}}
	\subfigure[Intermediate densities at the check node outputs.\label{fig:CheckDE_b}]{\includegraphics[width=0.48\textwidth,height=0.48\textheight,keepaspectratio]{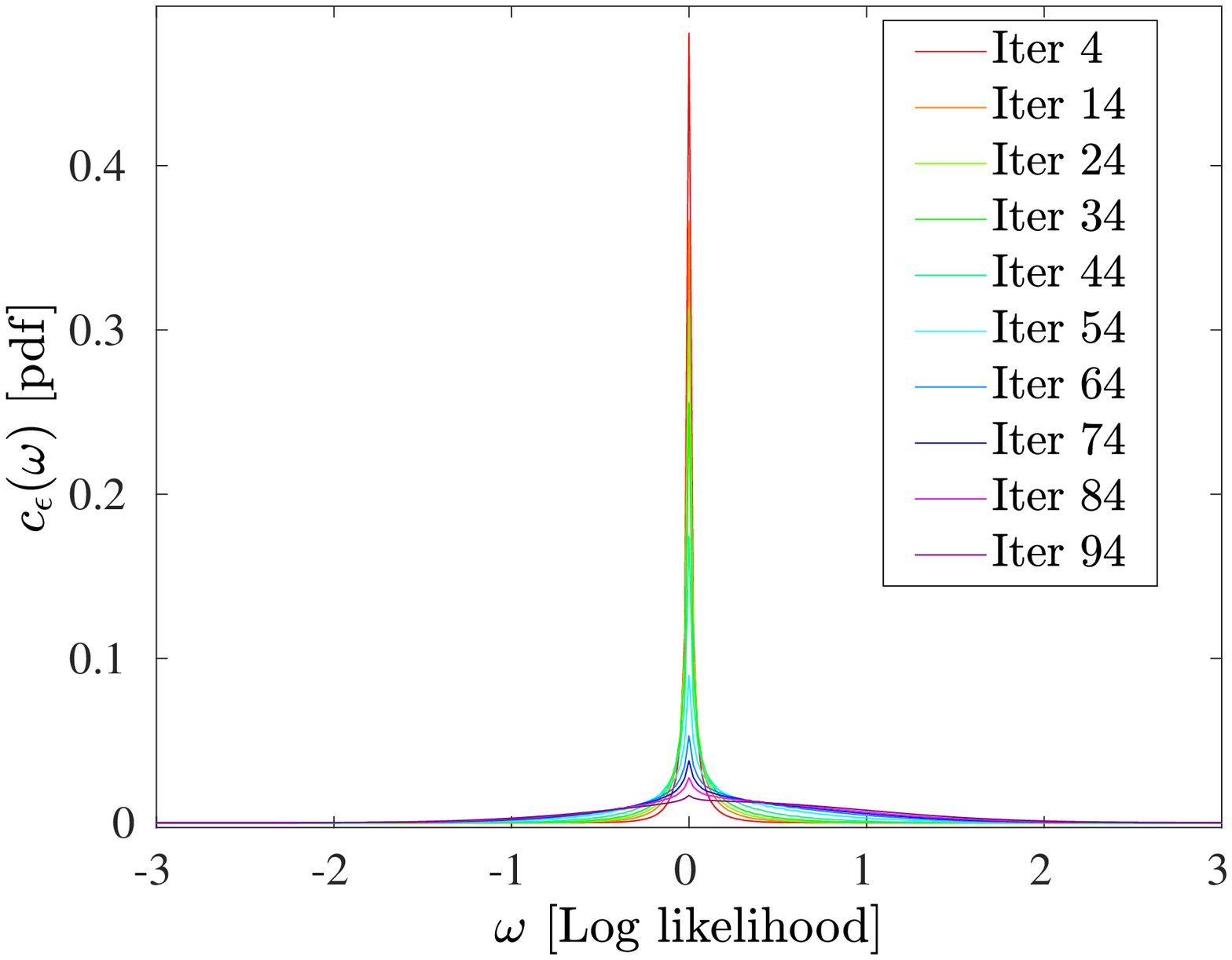}}\hfill
	\caption{The intermediate densities from iteration $4$ to iteration $94$, after each $10$ iterations for a code rate of $0.1$ and $E_b/N_0 = -1.0$ dB.}\label{fig:IntermediateDE} %Rate01DE2 %Rate01DECheck
\end{figure}

 For better understanding we plotted the evolution of the intermediate densities in the DE algorithm for the rate $0.1$ MET-LDPC code. As depicted in Fig. \ref{fig:IntermediateDE}, the Gaussian approximation is not valid for the check node output densities, but at the variable node outputs, the intermediate densities can be described by symmetric Gaussian distributions. Then in the process of the G-EXIT chart we can gradually change the mean of the Gaussian distribution from $0$ to $\infty$ \footnote{By $\infty$ we mean very large value for the LLR. We assume that in LLR domain $+25$ can be considered a large value and in this case we have almost zero error probability.}, and calculate the G-EXIT curves for the variable nodes and check nodes using~\eqref{eq:intermeditedensities}. %-\eqref{func: normalized dual G-EXIT no channel}.
\par
\begin{figure}[!ht]
	\begin{center}
	\includegraphics[width=0.6\textwidth,keepaspectratio]{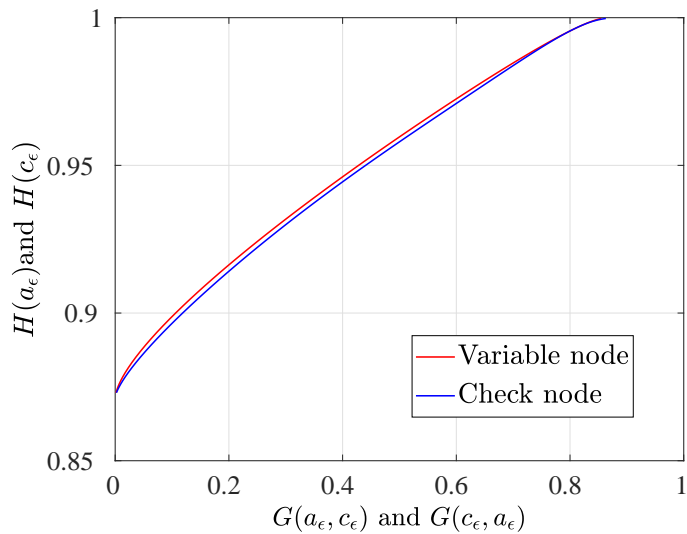}
	\end{center}\caption{The G-EXIT chart for the rate $0.1$ MET-LDPC code when the intermediate densities at the variable nodes are assumed to be symmetric Gaussian. The approximated threshold of the code is then equal to $-1.07$ dB.}\label{fig:G-EXITrate01_GA}
\end{figure}

 We also found the convergence threshold for the rate $0.1$ MET-LDPC code presented in Table~\ref{table:rate01urbanke} using our new proposed method. In comparison with DE the algorithm will find the convergence threshold very fast and the estimated threshold is $E_b/N_0 = -1.07$ dB ($\sigma^*_\textsubscript{App} = 2.53$). The threshold of the code given by DE was $-1.22$ dB ($\sigma^*_\textsubscript{DE} = 2.57$). The convergence behavior of this code is plotted in Fig. \ref{fig:G-EXITrate01_GA} when the intermediate densities at the variable nodes are assumed to be symmetric Gaussian.
\par Finally, to demonstrate the accuracy of this method, Table \ref{table:thresholdCompare} shows the convergence threshold for MET-LDPC codes taken from the literature and codes described in this paper using our G-EXIT method when Gaussian approximation is used in the variable node output ($\sigma^*_\textsubscript{App}$). The results are compared with the exact thresholds $\sigma^*_\textsubscript{DE}$ given by density evolution. In addition, to show the convergence speed we count the number of iterations for the DE to achieve an error probability smaller equal to $10^{-10}$. Furthermore, the asymptotic efficiency of each code is calculated as $R/C(s)$ at a SNR $s = 1/\sigma^2_\textsubscript{DE}$ where $R$ is the code rate and $C(s)$ is the capacity of the BIAWGN channel.
%$\sigma_\textsubscript{DE}, \sigma^*_{\scriptscriptstyle DE}$
\begin{table}[!ht]
	\caption{Comparison of the convergence threshold for different MET-LDPC codes using density evolution for BI-AWGN channel and our proposed approximation method. The $E_b/N_0~[dB]$ values correspond to the $\sigma^*_\textsubscript{DE}$. The error probability is set to be $10^{-10}$~.} % from Tables \ref{table:rate002designed} to \ref{table:rate05urbanke}
	\begin{center}\label{table:thresholdCompare}
	%\resizebox{0.95\hsize}{!}{
		\begin{tabular}{|c|c|c|c|c|c|c|c|}
			\hline
			$R^{ch}$ & Structure& $\sigma^*_\textsubscript{DE}$& $E_b/N_0 [dB]$& $\sigma^*_\textsubscript{App}$&$\sigma^*_\textsubscript{Sh}$&Iter.&Efficiency\\
			\hline
			$0.02$ &Sec. B of \cite{milicevic2017key}&$5.93$&$-1.48$&$5.88$&$5.96$&$688$&$98.87\,\%$\\
			\hline
			$0.02$ &Table~\ref{table:rate002designed}& $5.94$&$-1.50$&$5.82$&$5.96$&$467$&$99.28\,\%$\\
			\hline
			$0.05$ &Table~\ref{table:rate005urbankemodified}&$3.68$&$-1.32$&$3.65$&$3.73$&$387$&$97.36\,\%$\\
			\hline
			$0.10$ &Table~\ref{table:rate01urbanke}&$2.57$&$-1.22$& $2.53$&$2.59$&$263$&$ 98.61\,\%$\\
			\hline
			$0.50$ &Table~\ref{table:rate05urbanke}~\cite{richardson2002multi}&$0.965$&$0.305$&$0.96$&$0.9787$&$301$&$98.27\,\%$\\%95.09
			\hline
		\end{tabular}
		%}
	\end{center}
\end{table}
%%%%%%%%%%%%%%%%%%%%%%%%%%%%%%%%%%%%
%%%%%%%%%%%%%%%%%%%%%%%%%%%%%%%%%%%%
%%%%%%%%%%%%%%%%%%%%%%%%%%%%%%%%%%%%
\section{Conclusion}\label{sec:Conclusion}
We presented a practical approach to calculate theoretically optimal channel code rates of a multi-level coding/multi-stage decoding scheme suitable for reverse reconciliation in CV-QKD. Using $m$ individual levels of MLC-MSD, we can design a coding system for a specific SNR that operates close to the Shannon capacity.

Furthermore, we introduced the powerful tool and concept of G-EXIT charts for MET-LDPC codes and we proposed a new approximation method for the density evolution for such codes. For the codes used in this paper our DE approximation method predicts the threshold with an accuracy of $98\%$ to $99\%$. 

The presented methods can be used to quickly evaluate a given code which will enable a more time efficient design process without the need to run the density evolution algorithm to obtain the properties of a code, for instance the threshold. Using the methods we designed new MET-LDPC codes with rates $0.02,\,0.05$ and $0.1$ which outperform existing codes and which may be useful in existing CV-QKD implementations.
%%%%%%%%%%%%%%%%%%%%%%%%%%%%%%%%%%%%
%%%%%%%%%%%%%%%%%%%%%%%%%%%%%%%%%%%%
%%%%%%%%%%%%%%%%%%%%%%%%%%%%%%%%%%%%
\section*{Acknowledgment}
\includegraphics[width=0.05\textwidth,keepaspectratio]{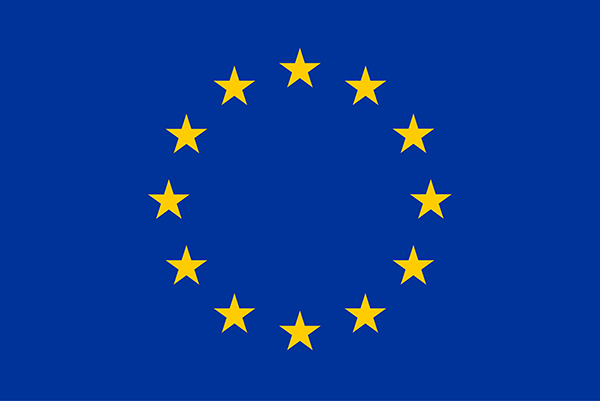}
This project has received funding from the European Union’s Horizon $2020$ research and innovation programme under grant agreement No $820466$ (CiViQ).
The authors thank the Quantum Innovation Center Qubiz funded by the Innovation Fund Denmark for support. HM, TG, and ULA acknowledge support from the Danish National Research Foundation, Center for Macroscopic Quantum States (bigQ, DNRF$142$).

%\bibliographystyle{IEEEtranTCOM}
%\bibliography{IEEEabrv,Bibliography}

\bibliographystyle{unsrt}
\bibliography{new} % file mwe.bib

\end{document}